# Widely tunable SNAP microresonators induced at the optical fiber intersections

M. Sumetsky
*Aston Institute of Photonic Technologies, Aston University, Birmingham B4 7ET, UK*
m.sumetsky@aston.ac.uk

## ABSTRACT

A widely tunable free spectral range (FSR) is essential for many microresonator applications, but achieving it in compact high-Q systems remains a significant challenge. Surface nanoscale axial photonics (SNAP) provides a route to such devices by localizing whispering-gallery modes through nanoscale effective-radius variations along an optical fiber. Recently, it has been experimentally demonstrated that side-coupling between two optical fibers can induce high-Q SNAP microresonators. In contrast to broadly explored monolithic microresonators, this configuration enables extensive mechanical tuning of the microresonator configuration and FSR through fiber bending, tilting, and twisting. Here, we develop the theory of SNAP microresonators induced at optical fiber intersections, which has remained largely unexplored. We consider weakly twisted fibers, whose geometry can be decomposed into tilting and bending. We show that extremely small fiber curvature critically affects the coupling-induced cutoff frequency variation, the shape of the effective axial potential, and the spectrum of the induced resonator. We discuss the physical origin of this curvature and show that accounting for it resolves the discrepancy between theory and experiment. These findings establish side-coupled optical fibers as a simple, reconfigurable platform for realizing microresonators with a widely tunable FSR.

## INTRODUCTION

Optical microresonators are fundamental components of modern photonic devices due to their compact size, high quality factors, and strong optical confinement. They have enabled numerous applications including optical filtering, sensing, nonlinear optics, frequency comb generation, optomechanics, and cavity quantum electrodynamics [1–5]. In many of these applications, including cavity electrodynamics [6-8], optical computing [9, 10], and optical frequency comb generation [11, 12], precise control of the resonator spectrum – particularly the free spectral range (FSR) – is of primary importance. However, in most monolithic whispering-gallery-mode (WGM) resonators such as microspheres, microtoroids, and microrings, the FSR is determined mainly by the resonator geometry and therefore remains essentially fixed after fabrication. Thermal, mechanical, and electro-optic tuning techniques can shift resonance frequencies but typically modify them simultaneously, leaving their spacing nearly unchanged [6, 13-16]. Achieving substantial and controllable FSR tunability in compact high-Q resonators, therefore, remains a significant challenge.

Surface nanoscale axial photonics (SNAP) provides a unique platform for the realization of high-Q microresonators based on nanoscale deformation of optical fibers [17]. In SNAP structures, WGMs, which circulate near the fiber surface and slowly propagate along the fiber axis, can be localized by nanoscale variation of the effective fiber radius. This variation is proportional to the variation of the WGM cutoff frequencies (CFs) $\omega_{m,q}^{(cut)}(z)$ along the fiber axis $z$, where $m$ and $q$ are the WGM azimuthal and radial quantum numbers. CFs act as *effective potentials* that confine WGMs along the fiber axis. Remarkably, variation in the effective radius on the nanometer scale can confine light along the fiber axis and form optical microresonators with high quality factors and unprecedented sub-angstrom fabrication precision [17-25].

Several techniques have been developed to create SNAP microresonators through controlled modification of optical fibers. Early demonstrations employed localized annealing of the fiber surface using a $CO_2$ laser, which relaxes frozen-in stresses and produces nanoscale radius variations along the fiber axis [17-21]. Additional fabrication methods, including nanoscale modification of microcapillary fiber surface by heated water [22], femtosecond-laser inscription [23] as well as flame and metal wire heating [24, 25] have also been used to introduce SNAP microresonators on optical fibers.

Recently, mechanical deformation of optical fibers was demonstrated as a particularly simple approach to creating tunable SNAP resonators. It was shown that strong bending of an optical fiber produces geometric deformation together with elasto-optic refractive-index variation, which generates an effective radius variation capable of localizing WGMs [26, 27]. This mechanism enables mechanically tunable microresonators whose spectral characteristics can be controlled by the fiber curvature. Controlled bending has been used to realize SNAP resonators with parabolic axial profiles and tunable picometer-scale FSR [27]. A miniature tunable optical delay line based on this mechanism has been demonstrated [28].

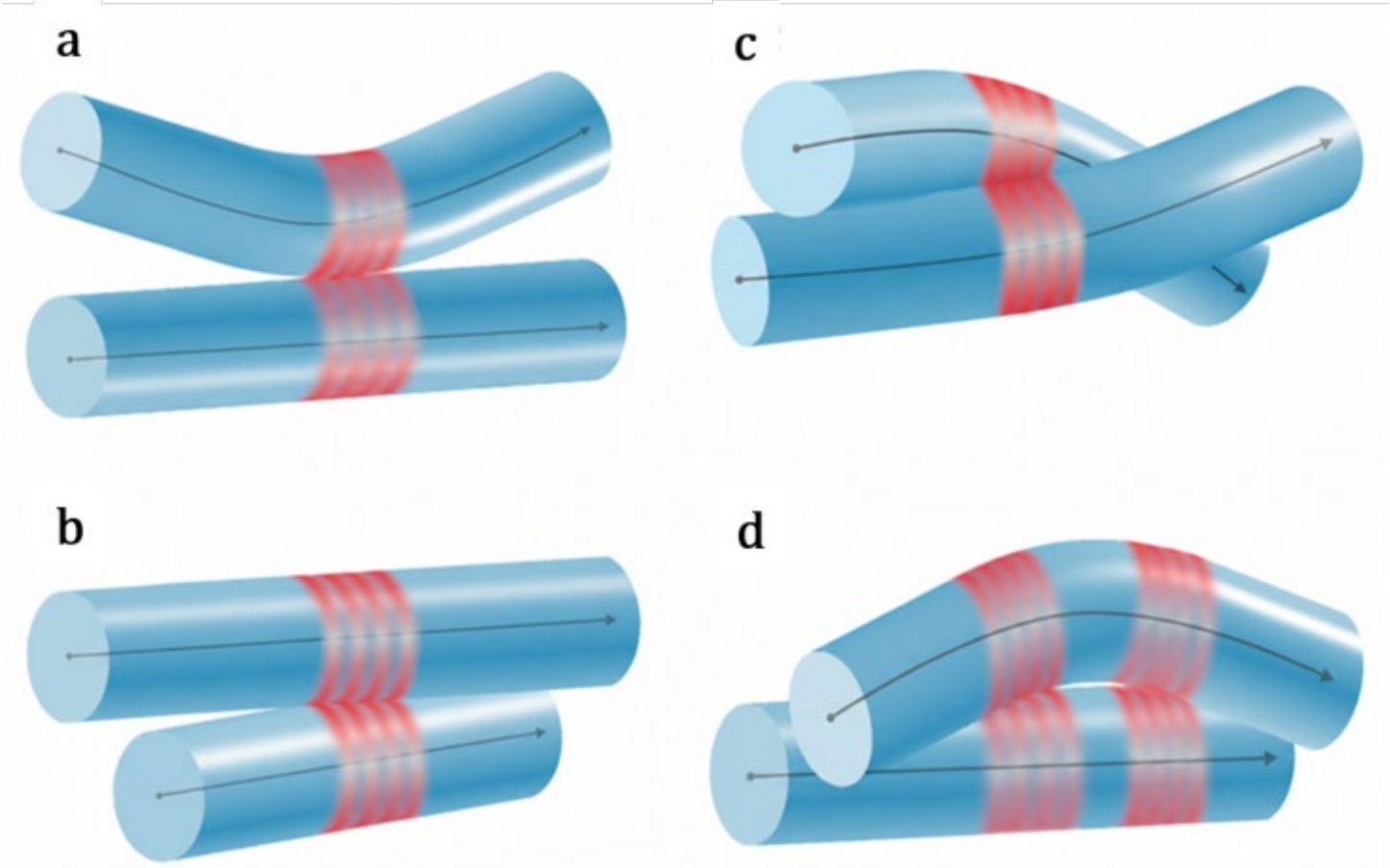

**Fig. 1**. a Bent, b tilted, and c twisted fiber configurations. d Bent fiber configuration with two contact points.

Another important mechanism for inducing SNAP resonators is based on coupling between nearby optical fibers (Fig. 1). It has been shown that side-coupled fibers can produce localized variations of the CF that form high-Q microresonators in the coupling region [29]. In this configuration, the resonator length and spectral characteristics can be tuned by adjusting the fiber curvature, enabling wide tuning of the microresonator FSR (Fig. 1a). Such mechanically reconfigurable resonators are promising for applications including tunable optical delay lines, frequency comb generators, and tunable miniature photonic circuits.

More recently, an even simpler configuration has been demonstrated experimentally: microresonators formed at the intersection of two straight optical fibers [30] (Fig. 1b). In this geometry, coupling between the fibers produces a localized variation of the CFs, which in turn localize WGMs slowly propagating along the fiber axis. Remarkably, extremely small relative tilting of the fiber segments by milliradian angles, which can be introduced by micrometer-scale translations of their ends, is sufficient to reshape this effective potential over millimeter-scale distances and produce substantial tuning of the resonator spectrum, including order-of-magnitude variation of the free spectral range [30]. This configuration is particularly attractive because it minimizes mechanical stress and enables compact implementations potentially compatible with MEMS technologies. Additionally, in Ref. [31] FSR tunable microresonators were demonstrated by parallel translation of optical fibers. The latter approach achieves better tuning precision at the expense of a larger translation range. Finally, preliminary experimental observations of microresonators induced in strongly twisted fiber configurations were reported in Ref. [32].

Despite the experimental demonstration of the FSR tunable microresonator configurations noted above [29-32], the physical mechanisms governing light localization in intersecting optical fibers and, in particular, the role of the geometric parameters of the considered fiber configurations were not well understood. Consequently, the correspondence of the experimental data with the simplified theory developed to date remained limited (see, e.g., Refs. [29, 30]).

In this work, we develop the theory of optical microresonators induced at the intersections of *weakly twisted* optical fibers, whose geometry can be decomposed into *tilting and bending*. The axial localization of WGMs in this configuration can be described by coupled wave equations governing the slow axial propagation of WGMs determined by the CF variation along the fiber axis. We show that the extremely small though unavoidable curvature of fibers plays a crucial role in determining the shape of induced microresonators and their spectra. Taking this curvature into account allows us to achieve good agreement between the developed theory and the experimental results.

## RESULTS

### Geometrical description of the twisted fiber configuration

We consider two twisted optical fibers with the same radius $r_0$ illustrated in Fig. 2 where we introduce Cartesian coordinates $(x, y, z)$. We choose the axis $z$ to be approximately aligned with the axes of fibers 1 and 2, and, consequently, axes $x$ and $y$ to be approximately in the transverse plane. The equations defining the axes of these fibers have the form $(x_j(z), y_j(z), z)$, $j = 1,2$. We are interested only in the small vicinity of the region where the fibers contact each other. In the vicinity of this region, the fiber axis profiles can be approximated by *quadratic dependencies on* $z$. In this approximation, both fiber axes have *zero torsion*, which is proportional to the third derivative, and therefore experience *planar bending* [33]. Under these assumptions, the equations describing the slightly curved fiber axes in the region of their proximity can be written as:

$$\begin{aligned} x_1(z) &= -r_0 - \frac{\Delta h_x}{2} + \frac{z^2}{2R_1} \\ y_1(z) &= -\frac{\alpha z}{2} \\ x_2(z) &= r_0 + \frac{\Delta h_x}{2} + \frac{z^2}{2R_2} \\ y_2(z) &= \frac{\alpha z}{2} \end{aligned} \tag{1}$$

Here $\Delta h_x > 0$, since the fibers cannot penetrate one another. Figs. 2(b), (c), and (d) show the mutually twisted fiber views along the plane normal to the axis $x$, $y$, and $z$, respectively. To keep the coupling length between the fibers much longer than their radius $r_0$, we assume that the tilt angle is small, $\alpha \ll 1$, (Fig. 2b) and the bending radii of fibers are much greater than the fiber radius, $|R_j| \gg r_0$ (Fig. 2c). More generally, we assume that, in the region of our interest, where the inter-fiber coupling may be significant, the deviation of the fiber axes in the coordinate plane $(x, y)$ is relatively small compared to the fiber radius $r_0$, i.e., $|x_j(z) - x_j(0)|, |y_j(z) - y_j(0)| \ll r_0$.

The distance between the fiber surfaces as a function of the coordinate $z$ can be expressed through the distance between fiber axes $d(z)$ as

$$\Delta d(z) = d(z) - 2r_0 = \sqrt{\left(x_1(z) - x_2(z)\right)^2 + \left(y_1(z) - y_2(z)\right)^2} - 2r_0 . \tag{2}$$

We are only interested here in the fiber inter-surface distance $\Delta d(z)$ which is small compared to the fiber radius $r_0$. Furthermore, to enable inter-fiber mode coupling, this distance should be of the order of a micron or less. Under these assumptions and using Eq. (1), we transform Eq. (2) to

$$\Delta d(z) = \Delta h_x + \frac{1}{2}\left(\frac{1}{R_{eff}} + \frac{\alpha^2}{2r_0}\right) z^2, \quad R_{eff} = \left(\frac{1}{R_2} - \frac{1}{R_1}\right)^{-1} . \tag{3}$$

where the effective fiber curvature radius $R_{eff}$ is introduced.

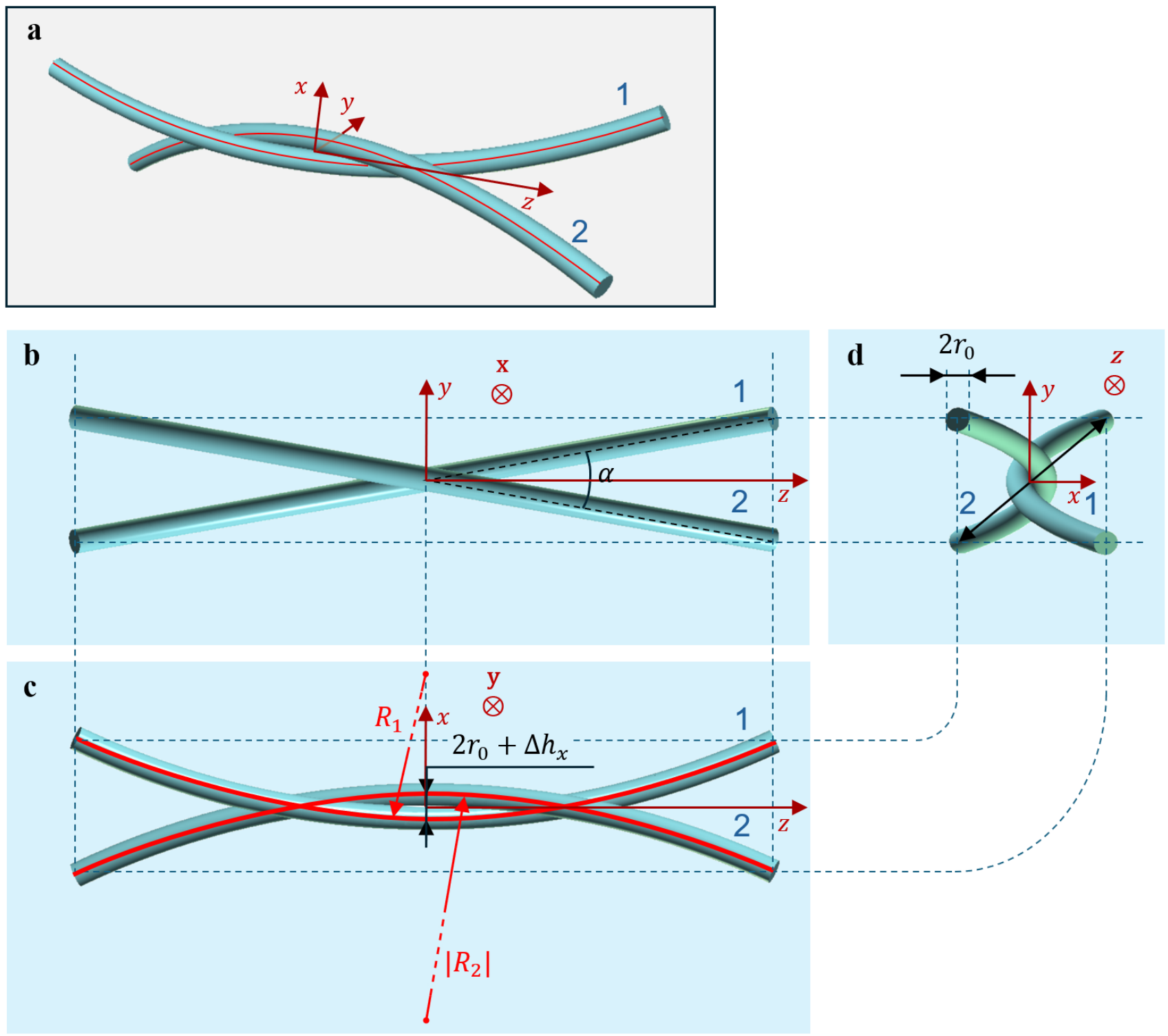

**Fig. 2.** Twisted fiber configuration. a 3D figure. b, c, and d Projections of figure (a) onto $(y, z)$, $(x, z)$, and $(x, y)$ planes, respectively.

The twisted fiber configuration shown in Fig. 2 reduces to the coplanar bent geometry illustrated in Fig. 1a [29] when the tilt angle vanishes ($\alpha = 0$), and to the tilted straight-fiber geometry illustrated in Fig. 1b [30] when the fiber curvatures vanish ($R_j = \infty$). As illustrated in Fig. 1, in the quadratic approximation considered the fibers may have one, two, or no contact points. In degenerate cases, they may contact over a curve that jointly belongs to both fibers. In this paper, we restrict our analysis to the case of a single minimum $\Delta h_x$ of the inter-fiber separation illustrated in Fig 2, when the case shown in Fig. 1d is excluded.

## Microresonators induced at fiber intersections: cutoff frequency resonance

The wave equation, which describes WGMs $E(\mathbf{r})$ *slowly propagating* along the configurations of fibers illustrated in Fig. 1 and described by Eqs. (1)-(3) can be written down as

$$\Delta E + \frac{\omega^2}{c^2} n^2(\mathbf{r}) E = 0 \tag{4}$$

Here $n(\mathbf{r})$ is the refractive index of the considered system, $c$ is the speed of light, and $\omega$ is the angular frequency of light. The WGMs in each of the *uncoupled* weakly bent fibers with the number $j = 1, 2$ can be presented as $E_j^{(0)}(\mathbf{r}) = \exp(i\beta_j(\omega)s_j)\exp(im_j\theta_j)Q_j(\rho_j)$, where $(\rho_j, \theta_j, s_j)$ is the local cylindrical coordinate system along the axis $s_j$ of fiber $j$. For the WGM frequencies close to the CFs $\omega_{c,j}$, the mode propagation constant as a function of light frequency $\omega$ can be expressed in each of the fibers as (see, e.g., [17])

$$\beta_j(\omega) = \frac{n_0}{c}\sqrt{2\omega_0\left(\omega - \omega_{c,j}\right)}, \tag{5}$$

where $n_0$ is the fiber refractive index and $\omega_0 \cong \omega_{cj}$. For a small coupling between fibers, the WGMs in each of them can be presented as $E_j(\mathbf{r}) = \Psi_j(z)\exp(im_j\theta_j)Q_j(\rho_j)$, where the dependence along the axis $z$ in fiber $j$ is described by functions $\Psi_j(z)$ satisfying the coupled wave equations (see Supporting Information (SI), Sections S1 and S2):

$$\begin{aligned} \frac{d^2\Psi_1}{dz^2} + \beta_1^2(\omega)\Psi_1 &= -\frac{\omega_0^2}{c^2} I_{12}(z)\Psi_2 \\ \frac{d^2\Psi_2}{dz^2} + \beta_2^2(\omega)\Psi_2 &= -\frac{\omega_0^2}{c^2} I_{21}(z)\Psi_1 \end{aligned} \tag{6}$$

where $I_{ij}$ are the coupling coefficients. For the twisted fiber configuration considered, we find (see Sections S1 and S2 of SI):

$$\begin{gathered} I_{12}(z) = I_{21}^*(z) = I_0 \exp\left[i\beta_0 z - \sigma d(z)\right], \\ I_0 \simeq \frac{2\pi^{-1/2}(-1)^q n_0^2}{(n_0^2-1)^{3/4}}\left(\frac{c}{\omega_0 r_0}\right)^{3/2}, \quad \beta_0 = \frac{2\pi\alpha n_0}{\lambda_0}, \quad \sigma = \frac{\omega_0(n_0^2-1)^{1/2}}{c}, \end{gathered} \tag{7}$$

where $q$ is an integer equal to the sum of the radial quantum numbers of fiber modes and the effective fiber curvature radius $R_{eff}$ and separation between fibers $\Delta d(z)$ are defined in Eq. (3). The flux conservation law, which simply follows from Eqs. (6) and (7), shows that, in the approximation considered, the coupling-induced losses are absent.

From Eqs. (6) and (7), the spatial variation of the coupling coefficients $I_{ij}$ is determined by the characteristic distance $\beta_0^{-1}$ inverse proportional to the tilt angle $\alpha$ and characteristic distance $\sigma^{-1}$. As expected, the coupling coefficients $I_{ij}$ exponentially decrease with the separation between fibers $\Delta d(z)$. We find from Eq. (3) for $\Delta d(z)$ that for the positive $R_{eff}$, the fiber curvatures $1/R_j$ and tilt angle $\alpha$ are jointly responsible for the localization of coupling and for reducing the induced microresonator's axial size. However, for negative $R_{eff}$, the curvature compensates for the tilt at $R_{eff} = -2r_0\alpha^{-2}$. The latter relation between the effective radius and tilt angle is plotted for $r_0 = 20\ \mu$m, $r_0 = 62.5\ \mu$m, and $r_0 = 100\ \mu$m in Fig. 3. For $R_{eff} < 0$ and $|R_{eff}| < 2r_0\alpha^{-2}$, the single-point inter-fiber contact configuration can evolve into two inter-fiber point contacts or a contact along a continuous line segment joining the fiber surfaces whose length can be enhanced by the Van der Waals attraction [30, 34, 35].

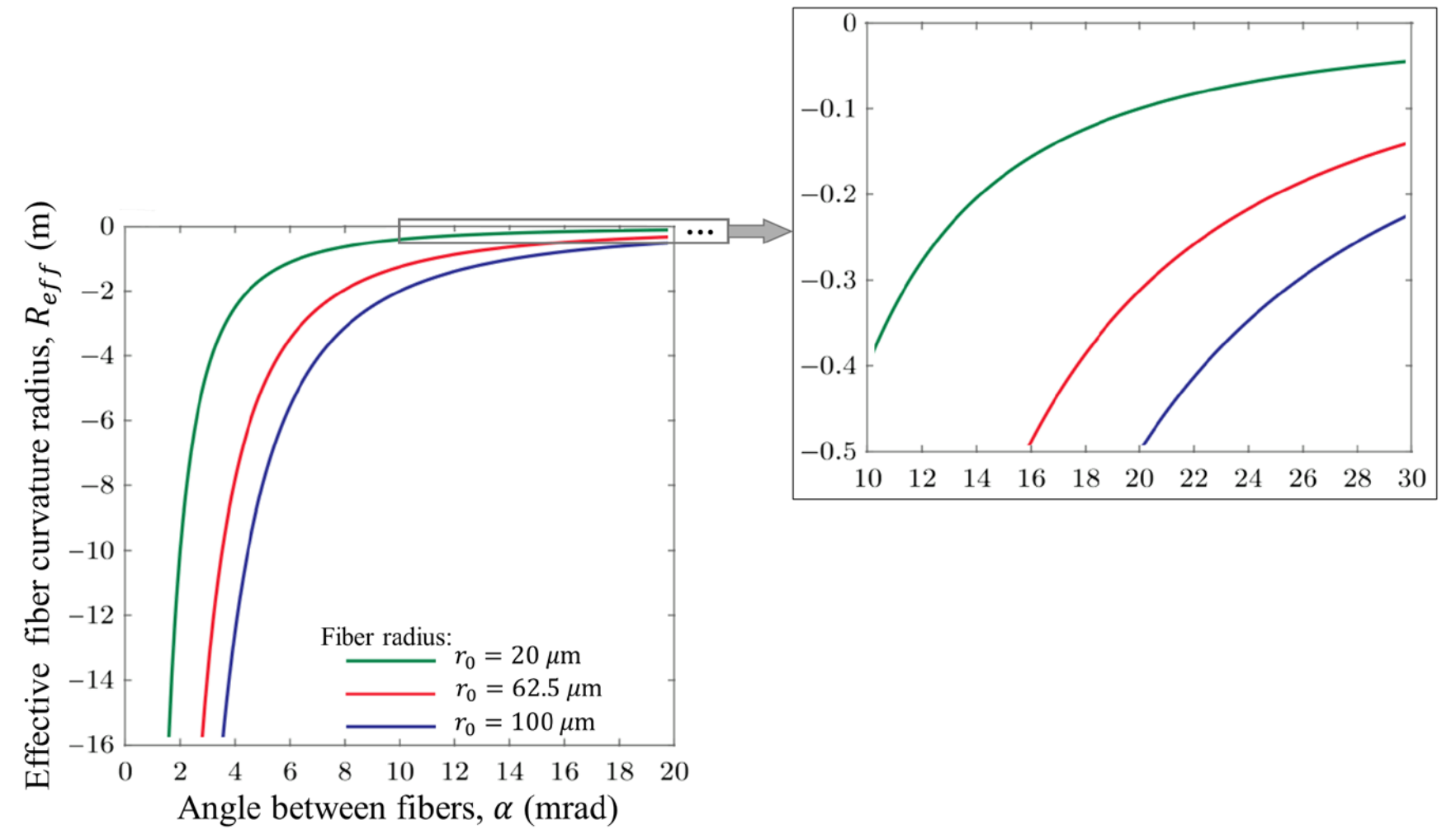

**Fig. 3.** The relation between the tilt angle $\alpha$ and the negative effective curvature radius $R_{eff} = -2r_0\alpha^{-2}$ compensating for this tilt for $r_0 = 20$ μm (green curve), $r_0 = 62.5$ μm (red curve), and $r_0 = 100$ μm (blue curve).

It is shown in Section S3 of SI that CFs $\omega_c^{\pm}(z)$ of the coupled fiber system considered can be found in the WKB approximation as

$$\omega_c^{\pm}(z) = \frac{1}{2}\left(\omega_{c,1} + \omega_{c,2} \pm \sqrt{\left(\omega_{c,1} - \omega_{c,2}\right)^2 + \frac{\omega_0^2}{n_0^4} I_{12}(z) I_{21}(z)}\right) \tag{8}$$

The characteristic dependencies of $\omega_c^{\pm}(z)$ near the coupling region are illustrated in Fig. 4, which exhibits the well-known effect of coupling-induced level repulsion [36]. As a result of repulsion, the larger CF, $\omega_c^{+}(z)$, is bent up and forms a potential barrier, while the smaller one, $\omega_c^{-}(z)$, is bent down and forms a quantum well, i.e., a microresonator induced by fiber coupling. As shown in Section S3 of SI, the propagation constants of the system (now depending on the axial coordinate $z$) can be expressed through the CFs $\omega_c^{\pm}(z)$ similar to Eq. (5) as

$$\beta_{\pm}^2(z,\omega) = 2\omega_0 \frac{n_0^2}{c^2}\left(\omega - \omega_c^{\mp}(z)\right). \tag{9}$$

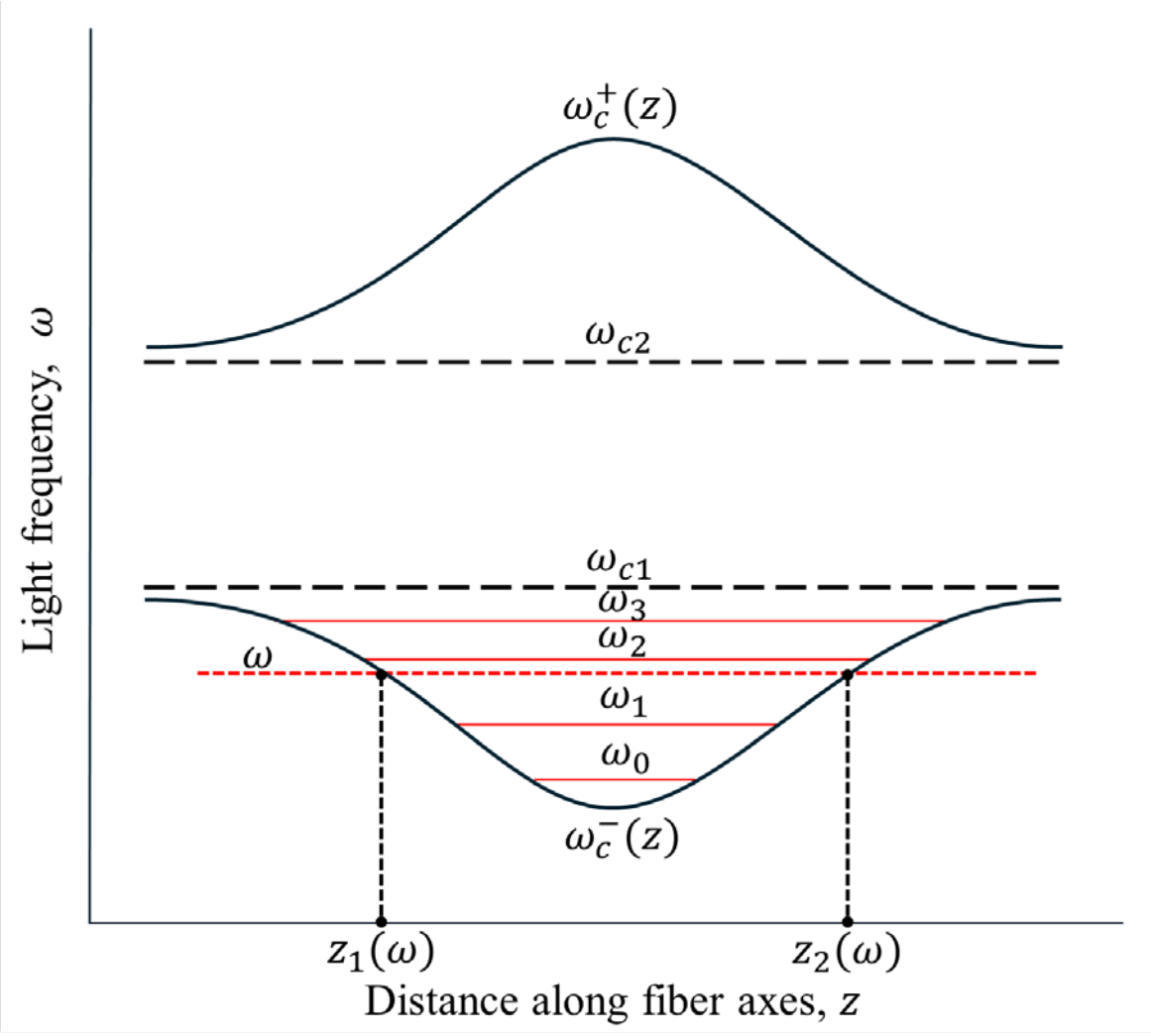

**Fig. 4.** Behavior of CFs $\omega_c^{\pm}(z)$ near the coupling region.

In the WKB approximation, for a known CF quantum well profile, $\omega_c^-(z)$, the resonator eigenfrequencies can be found from the Bohr-Sommerfeld quantization rule [36],

$$\int_{z_1(\omega_k)}^{z_2(\omega_k)} \beta_+(z,\omega_k)dz = \pi\left(k+\frac{1}{2}\right), \quad k \gg 1, \text{ integer.} \tag{10}$$

Here $z_j(\omega)$ are the WGM turning points (zeros of $\beta_+(z,\omega)$) shown in Fig. 4.

The asymptotic expression for the coupling coefficients $I_{ij}$ in Eq. (7) loses accuracy at small fiber separations, where the coefficients are largest and have the greatest effect on the CFs. Therefore, in Section S4 of SI we calculate these coefficients numerically using the multipole method. In the adiabatic approximation, which is valid for small tilt angles $\alpha$ and large bending radii $R_{eff}$ considered here, we find the local CFs as eigenfrequencies of the two-dimensional coupled-fiber cross-section, calculated as functions of the fiber separation $\Delta d(z) = \Delta h_x + \frac{1}{2}\left(R_{eff}^{-1} + \alpha^2/2r_0\right)z^2$ (see Eq. (3)). The dependence of $I_{12}$ on $\alpha$, $R_{eff}$, and $z$ is incorporated in $I_{12}$ through its dependence on $\Delta d(z)$. At the exact matching of CFs of uncoupled fibers, $\omega_{c,1} = \omega_{c,2}$, the CF quantum well profile is found from Eq. (8) as $\omega_c^-(z) = \omega_{c,1} - \frac{1}{2}\omega_0 n_0^{-2}|I_{12}(\Delta d(z)|$. Similarly, for the coupling much smaller than the CF separation, $I_0 \ll 2{n_0}^2\left|\omega_{c,2} - \omega_{c,1}\right|/\omega_0$, we arrive at the CF quantum well whose FWHM is $\sqrt{2}$ times smaller than that for the exact CF matching: $\omega_c^-(z) = \omega_{c,1} - \frac{1}{4}\omega_0^2 n_0^{-4}\left|\omega_{c,2} - \omega_{c,1}\right|^{-1}|I_{12}(\Delta d(z)|^2$. For small fiber separation, when $\Delta d(z) \lesssim c\omega_0^{-1}(n_0^2 - 1)^{-\frac{1}{2}} \cong 0.2$ µm for $\omega_0 = 2\pi \cdot 193$ THz and $n_0 = 1.44$, these approximations lose accuracy and have to be corrected by numerical calculations (see Figs. S6 and S7 of SI).

Comparison of the numerical calculation and analytical approximations of the CF splitting based on Eq. (7) on Fig. S6 of SI shows that their relative difference is less than 30% for $\Delta d(z) < 1$ µm achieving zero at $\Delta d(z) \cong 0.29$ µm. Therefore, it is reasonable to use the analytical results of this section to find the estimates of the axial FWHM $\Delta z_{FWHM}$ of the induced resonator and also of the FSR $\Delta\omega_{FSR}$ near its bottom where the CF quantum well is close to parabolic. At the exact resonance, $\omega_{c,2} = \omega_{c,1}$, we find from Eqs. (7) and (8):

$$\Delta z_{\mathrm{FWHM}} = 4\left[\frac{\ln(2)cr_0}{\left({n_0}^2-1\right)^{1/2}\omega_0\left(\dfrac{2r_0}{R_{\mathrm{eff}}}+\alpha^2\right)}\right]^{1/2}, \quad \Delta\omega_{\mathrm{FSR}} = \frac{c^{5/4}\left(\dfrac{2r_0}{R_{\mathrm{eff}}}+\alpha^2\right)^{1/2}\exp\left[-\dfrac{\omega_0}{2c}\left({n_0}^2-1\right)^{1/2}\Delta h_x\right]}{2^{1/2}\pi^{1/4}n_0\left({n_0}^2-1\right)^{1/8}{\omega_0}^{1/4}{r_0}^{5/4}}. \tag{11}$$

Based on Eq. (11), Fig. 5 illustrates how fiber tilting and bending jointly control the axial width and FSR of the induced microresonator. In this figure we assume $\Delta h_x = 0$ and exact CF resonance, when the depth of the CF quantum well calculated from Eqs.(7) and (8) is 25. 5 GHz compared to 19.7 GHz numerically calculated in SI. Increasing $\alpha$ narrows the resonator (Fig. 5a) and increases its FSR (Fig. 5b). At small tilt angles, fiber curvature determines the resonator width and a nonzero FSR. As $\alpha^2$ becomes much larger than $2r_0/R_{\mathrm{eff}}$, tilting dominates and the curves for different curvature radii approach one another. In this regime, the FWHM decreases approximately as $1/\alpha$, whereas the FSR increases approximately linearly with $\alpha$. Consequently, weakly curved fibers allow a broader relative tuning range, while stronger bending reduces the sensitivity to small variations of the tilt angle. Increasing $\Delta h_x$ to 0.2 µm, when, as noted above, the considered approximation becomes much more accurate, we find that the dependence of FWHM $\Delta z_{FWHM}$ does not change, while the FSR $\Delta\omega_{FSR}$ becomes 1.5 times smaller.

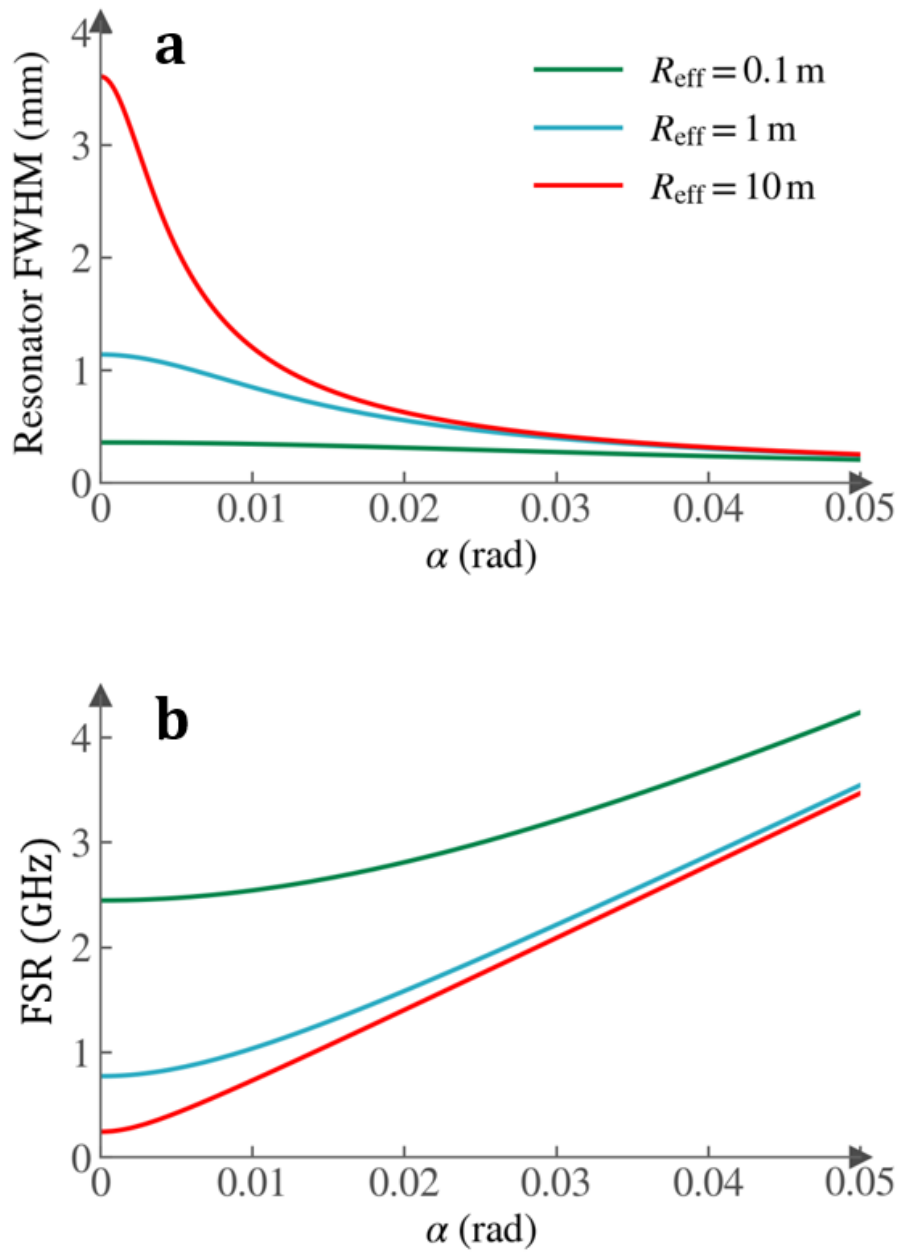

**Fig. 5.** a Microresonator FWHM $\Delta z_{\rm FWHM}$ and b FSR $\Delta\omega_{\rm FSR}/(2\pi)$ as functions of the tilt angle $\alpha$ for effective curvature radii $R_{\rm eff} = 0.1$ m, 1 m, and 10 m. The curves are calculated from Eq. (11) with $r_0 = 62.5$ µm, $n_0 = 1.45$, and $\omega_0/(2\pi) \approx 193.4$ THz.

Using Eq. (11), we can estimate the achievable precision of experimental FSR tuning. For the parameters of Fig. 5b, a rotation stage with an angular repeatability of 0.1 mrad limits the tilt-induced FSR error to less than 10 MHz, provided that fiber curvature and contact conditions remain unchanged. Equivalent angular control can be obtained by lateral translation with 1 µm repeatability over a 10 mm lever arm. These requirements are readily achievable with precision stages. The FSR can also be tuned by adjusting the minimum inter-fiber gap $\Delta h_x$. For the parameters of Fig. 5b, increasing this gap by 0.1 µm reduces the predicted FSR by approximately 20%. With stable fiber mounting, the 10 nm positioning repeatability available with closed-loop piezoelectric stages would correspond to approximately 2% FSR repeatability, or about 20 MHz for a 1 GHz FSR, provided that the fiber curvature and tilt remain unchanged. The silica thermal response shifts individual resonance frequencies by order 1 GHz/K near 1550 nm. Therefore, limiting this drift to 0.1 GHz requires temperature stability of approximately 0.1 K. A uniform temperature change shifts neighboring resonances together, with a much smaller effect on the FSR. Finally, fiber segments whose CFs match to within 0.1 GHz can be selected from the same parent fiber [37].

## Microresonators induced at fiber intersections: cutoff frequency perturbation

In the previous Section, we considered resonant coupling between fibers with two closely spaced CFs, separated from all other CFs. Then, the CF shift due to coupling can be comparable to the separation between them. Now, we assume the case where the CFs of fibers are perturbed by their side-coupling so that this shift is much smaller than the CF separation. Then, the variation of CFs is determined by the coupling of multiple transverse eigenstates of fibers. For weak inter-fiber coupling, the WGMs in the fiber $j$ can be presented as a linear combination $\sum_{p_j=1}^{P_j} Q_{j,p}(\rho_j)\exp\left(im_{j,p_j}\theta_j\right)\Psi_{j,p_j}(z)$ where $\Psi_{j,p_j}(z)$ are determined by the coupled wave equations of $P_j$ fiber modes, which are the direct generalization of Eq. (6). For the uncoupled fiber $j$, we have $\Psi_{j,p_j}(z)\sim\exp(i\beta_{j,p_j}(\omega)z)$, where $\beta_{j,p_j}(\omega) = (n_0/c)(2\omega(\omega-\omega_{c,j,p_j}))^{1/2}$ and $\omega_{cj,p_j}$ is a CF. (see Eq. (9)). In the adiabatic approximation for coupled fibers, we set $\Psi_{j,p_j}(z)\sim A_{j,p_j}\sum_\pm \exp\left(\pm i\int^z \beta(z',\omega)dz'\right)$, where $\beta(z,\omega) = (n_0/c)(2\omega(\omega-\omega_c(z)))^{1/2}$ and $\omega_c(z)$ is the CF of the coupled fibers. Then, the coupled wave equations are transferred to the system of linear algebraic equations

$$\frac{2n_0^2}{\omega}\left(\omega_c(z)-\omega_{cj,p_j}\right)A_{j,p_j}+\sum_{p_{3-j}} I_{j,p_j,p_{3-j}}(z)A_{3-j,p_{3-j}}=0,\quad j=1,2,\quad p_j=1,2,\ldots,P_j, \tag{12}$$

where $I_{j,p,q}$ are the coupling coefficients. For a small inter-fiber coupling, when $\omega n_0^{-2}\left|I_{j,p_j,p_{3-j}}\right|$ is much smaller than the separation between the CFs of uncoupled fibers, $|\omega_{c,j,p_j}-\omega_{c,3-j,p_{3-j}}|$, the field amplitudes $A_{j,p_j}$ can be found by the perturbation theory. In particular, $\omega_c(z)$ is found by setting the determinant of this equation equal to zero, leading to the algebraic equation of the $P_1+P_2$-order for $\omega_c(z)$. Assuming that one of the solutions, $\omega_c^{(pj)}(z)$, is close to the unperturbed eigenfrequency $\omega_{c,j,p_j}$, we find:

$$\omega_c^{(p_j)}(z)=\omega_{c,j,p_j}+\frac{\omega_0^2}{4n_0^4}\sum_{p_{3-j}=1}^{P_{3-j}}\frac{I_{j,p_j,p_{3-j}}(z)I_{3-j,p_{3-j},p_j}(z)}{\omega_{c,j,p_j}-\omega_{c,3-j,p_{3-j}}}. \tag{13}$$

For the simplest case $P_1=P_2=1$, this result coincides with Eq. (8) under the condition of weak coupling, $|\omega_{c,1}-\omega_{c,2}|\gg\omega n_0^{-2}|I_{12}I_{21}|^{1/2}$. In the approximation considered, the absolute values of the products of coupling coefficients $I_{j,p_j,p_{3-j}}(z)I_{3-j,p_{3-j},p_j}(z)$ in Eq. (13) coincide with that defined previously, $I_{12}(z)I_{21}(z)$, and are independent of quantum numbers of solutions (see Eq. (7)). Therefore, Eq. (13) can be rewritten as

$$\begin{gathered}\omega_c^{(p_j)}(z)=\omega_{c,j,p_j}+\Delta\omega_{p_j}\exp\left(-2\sigma\Delta d(z)\right),\\ \Delta\omega_{p_j}=\frac{1}{\pi(n_0^2-1)^{3/2}\omega_0}\left(\frac{c}{r_0}\right)^3\sum_{p_{3-j}=1}^{P_{3-j}}\frac{1}{\omega_{c,j,p_j}-\omega_{c,3-j,p_{3-j}}},\quad \sigma=\frac{\omega_0(n_0^2-1)^{1/2}}{c},\end{gathered} \tag{14}$$

where $R_{eff}$ and $\Delta d(z)$ are defined in Eq. (3). Critically, this equation demonstrates the *universal Gaussian axial dependence* of the CFs $\sim\exp\left[-\sigma\left(R_{eff}^{-1}+\alpha^2/2r_0\right)z^2\right]$, i.e., simply expressed through the geometric parameters of the considered fiber configuration and *independent of the quantum numbers* $p_j$ of coupled transverse eigenstates $Q_{j,p_j}(\rho_j)$. The sum in Eq. (14) depends on the values of the fiber CFs, includes both positive and negative terms, and, thus, can be both positive and negative. A positive shift $\Delta\omega_{p_j}$ forms a CF barrier, whereas a negative shift forms a CF well. The order of magnitude of this sum can be estimated as $1/\Delta\omega_{FSR}$ or greater where $\Delta\omega_{FSR}$ is the characteristic free spectral range of the fiber CFs. Consequently, the preexponent is estimated as $\Delta\omega_{p_j}\sim(c/r_0)^3(\omega_0\Delta\omega_{FSR})^{-1}$. For example, setting $\omega_0=2\pi\cdot200$ THz, $r_0=60$ µm, and $n_0=1.45$, we find $\Delta\omega_{FSR}=c/(n_0r_0)=2\pi\cdot0.55$ THz, and $\Delta\omega_{p_j}\sim\pm2\pi\cdot1$ GHz. Remarkably, the order of this CF variation coincides with that observed for the SNAP microresonators [17-25]. Rescaling $\Delta\omega_{p_j}$ to the effective radius variation $\Delta r_{eff}=-r_0\Delta\omega_{p_j}/\omega_0$ commonly used in the description of SNAP microresonators, we arrive at the nanoscale $\Delta r_{eff}\sim1$ nm or greater, characteristic for SNAP structures.

As we show in Sections S4 and S5 of SI, even nanometer-scale differences between fiber radii can substantially alter the structure of their WGM CF spectra and the inter-fiber mode coupling. To demonstrate this sensitivity, in Fig. S4 of SI we compare the spectra of silica fibers with initial radii $r_0=62.5$ µm and $\omega_0\cong2\pi\cdot196$ THz introducing their radius differences of only 10-30 nm. These nanoscale geometrical changes shift corresponding CFs by approximately 30–100 GHz lifting the degeneracy of modes with the same azimuthal and radial quantum numbers. In Fig. S5 of SI, we examine how individual WGMs of fiber 2 contribute to the CF shift of a selected fundamental radial WGM of fiber 1. We show that, despite the dense transverse spectrum illustrated in Fig. S4, only a limited subset of modes contributes appreciably. For identical fibers, resonant coupling between matching modes dominates at larger separations $\Delta d(z)$, while other modes become increasingly important as the separation decreases. A small radius mismatch detunes the matching modes and increases the relative importance of these additional contributions. The increasing contribution of other modes at smaller $\Delta d(z)$, together with the detuning caused by radius mismatch, limits the accuracy of the analytical approximation described by Eqs. (7) and (14). However, in the adiabatic approximation, which is justified for sufficiently small tilt angle $\alpha$ and effective fiber curvature $R_{eff}^{-1}$ or interest here, the numerical multipole method described in Section S4 of SI can be applied.

## Experimental considerations

Recent experimental demonstration of microresonators created at the crossing of two straight fibers showed that the variation of a tilt angle by a *milliradian* order introduces the variation of the microresonator's length by a *millimeter* order [30]. In the theoretical estimates of Ref. [30], the crossing fiber segments were assumed to be ideally straight, while the correspondence with the experimental data was insufficient. Here, we suggest that the observed discrepancy between theory and experiment may partly reflect neglect of very small but finite fiber curvatures. The estimates below test the plausibility of this interpretation. We show that a physically justified meter-order-long curvature radius in millimeter-order-long fibers can dramatically change the inter-fiber coupling length and the microresonator spectrum.

Optical fiber segments, even very short ones, are never ideally straight. They can be divided into those with permanent and those with temporary curvature. Temporary curvature can be introduced mechanically or by local heating, while permanent curvature can be introduced by fiber melting or annealing. As an example of temporary induced curvature, Fig. 6a shows a fiber segment bent by its own weight. We assume that the fiber segment has a free end and is supported at a point away from its end by a distance $L$. Then, following the Euler-Bernoulli beam theory [38], the fiber curvature radius at the support point can be found as

$$R_{weight} = \frac{Er_0^2}{2\rho g L^2}, \tag{15}$$

where $E$ is the Young's modulus, $\rho$ is the fiber density, and $g$ is the acceleration due to gravity. For a silica fiber considered, we set $E = 72 \cdot 10^9$ N/m$^2$, $\rho = 2200$ kg/m$^3$, $g = 9.81$ m/s$^2$, and $r_0 = 62.5$ µm. Then, for $L = 2$ cm, we find $R_{weight} = 16$ m and for $L = 5$ cm, we find $R_{weight} = 2.6$ m.

A temporary induced curvature can also be induced by improper alignment of the fiber's fixed ends, leading to a deviation $\delta$ of the fiber end from its direction at the fiber intersection point (Fig. 6b). The *smallest* deviation $\delta_{min}$ corresponds to the ideal situation in which the fiber orientation at the fiber intersection point is not altered by misalignment. The corresponding fiber curvature radius is related to this deviation as

$$R_{mis} = \frac{L^2}{3\delta_{\min}} \tag{16}$$

For a characteristic misalignment of translation stages $\delta_{min} \sim 100$ µm and $L \sim 2$ cm we find $R_{mis} \sim 1$ m.

A commonly used approach to clean the fiber surface is heating it with a $CO_2$ laser or flame. This process is similar to fiber annealing and may cause asymmetric (along the fiber cross-section) release of frozen-in stresses, leading to permanent fiber bending. The release of residual stresses, whose characteristic value for silica fibers is $\sigma \sim 10^7$ N/m$^2$ [39, 40] leads to fiber shrinkage $\Delta\varepsilon = \sigma/E \sim 10^{-4}$. For side annealing, we determine the smallest possible curvature radius by modeling the fiber as a bi-material beam composed of materials with equal widths and Young moduli, and assuming that stresses are relaxed in one material and remain unchanged in the other. Then the *smallest* induced fiber curvature radius $R_{anneal}$ can be estimated from the Timoshenko equation [41] as

$$R_{anneal} = \frac{4r_0}{3\Delta\varepsilon} = \frac{4r_0 E}{3\sigma} \sim 0.6 \text{ m}. \tag{17}$$

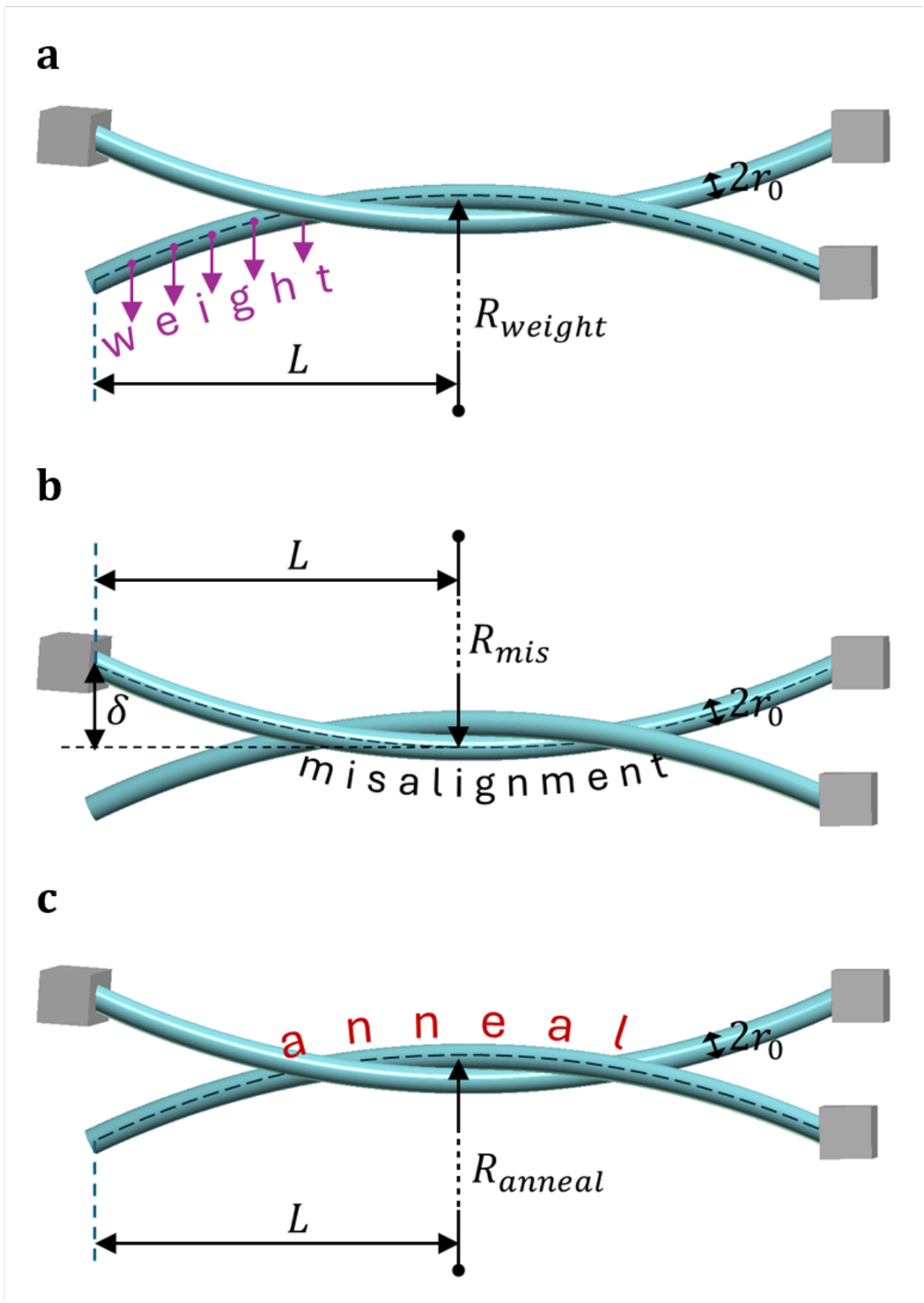

**Fig. 6.** a A fiber segment bent by its own weight. b A fiber segment bent by misalignment. c A fiber segment bent by side annealing.

It follows from the estimates based on Eqs. (15)-(17) that, for different reasons, the fiber segments of the considered system can have a *meter-order* curvature radius. These dramatically small bending effects have to be taken into account in the explanation of the results of recent experiments [30]. In Ref. [30], SNAP microresonators were created at the intersection of optical fiber segments, which were assumed to be straight, i.e., to have $R_1 = R_2 = \infty$. The FSR of these resonators was tuned by the milliradian-scale variation of the angle $\alpha$ between fibers. The microresonators were characterized using spectrograms, i.e., 2D plots of their spectra measured with a transverse microfiber taper connected to the optical spectrum analyzer. The microfiber was scanned along the length of one of the fibers, periodically touching its surface at a 2 μm resolution. The spectrograms of two of these microresonators corresponding to the angles $\alpha_1 = 0.01$ rad and $\alpha_2 = 0.005$ rad are shown in Figs. 7a and 7b, respectively. In these spectrograms, the wavelength dependence along the vertical axes used in [30] was rescaled to the frequency dependence aligned with the present paper. Generally, the coupling-induced frequency splitting and shifting depend on the tilt angle $\alpha$ and effective fiber curvature radius $R_{eff}$ only through the fiber separation $\Delta d(z)$ defined by Eq. (3). Therefore, for $\Delta h_x = 0$, we have

$$\omega_c^{(p_j)}(z) = \omega_{c,j,p_j}(\Delta d(z)) = \omega_{c,j,p_j}\left(\frac{1}{2}\left(\frac{1}{R_{eff}} + \frac{\alpha^2}{2r_0}\right)z^2\right). \quad (18)$$

Consequently, fiber configurations having the same value of $\left(R_{eff}^{-1} + \alpha^2/2r_0\right)$ should exhibit identical WGM spectra.

If the fibers in Ref. [30] were straight, i.e. $R_1 = R_2 = \infty$, Eq. (18) would predict that the CF profile in Fig. 7b could be mapped onto that in Fig. 7a by rescaling the distance from the contact point, $z \to \chi z$, with the factor $\chi = \alpha_2/\alpha_1$. However, matching the measured profiles requires $\chi = 0.74$, rather than the measured angle ratio $\alpha_2/\alpha_1 = 0.5$. If instead we assume a finite effective fiber curvature radius $R_{eff}$ that is the same for both configurations, Eq. (18) gives

$$\chi = \left( \frac{2r_0 + \alpha_2^2 R_{eff}}{2r_0 + \alpha_1^2 R_{eff}} \right)^{1/2} \tag{19}$$

Using the measured scaling factor $\chi = 0.74$ and $r_0 = 62.5$ µm, Eq. (19) gives $R_{eff} \approx 1.9$ m.

A quantitative numerical calculation of the CF profiles for the configuration in Ref. [30] requires the values of uncoupled CFs of the participating WGMs, which are not known with sufficient precision and are sensitive to nanoscale variations in the fiber radii (see Figs. S4 and S5 of SI). Therefore, it is also interesting to compare the experiments with the analytical models described above. For this purpose, we assume the Gaussian axial dependence $\omega_c(z) = \omega_{c1} + \Delta\omega \exp(-2\sigma z^2)$, obtained from perturbation theory and Eq. (14), and fit the spectrograms in Figs. 7a and 7b by adjusting the effective curvature radius $R_{eff}$ and the signed CF shift $\Delta\omega$ while retaining the experimentally measured angle $\alpha$. Comparison of the experimental spectrogram in Fig. 7a with the calculated CF profile and axial spectrum in Fig. 7c shows good agreement for $\alpha = 0.01$ and $R_{eff} = 0.932$ m, or, equivalently within this model, for $\alpha = 0.0153$ and $R_{eff} = \infty$. For comparison, Fig. 7d shows the calculation for the measured angle $\alpha = 0.01$ with straight fibers, $R_{eff} = \infty$. The numerical examples in Section S5 of SI indicate a weaker dependence of the CF shift on fiber separation than predicted by the perturbative planar approximation. For straight fibers, reproducing this behavior with the analytical profile would require a reduction in the effective value of $\alpha$, whereas fitting the experimental spectrum requires an increase in $\alpha$ above its measured value. Thus, the discrepancy is not explained by this correction to the coupling approximation alone and motivates including finite fiber curvature. Similarly, the experimental spectrogram in Fig. 7b, measured at $\alpha = 0.005$, is in reasonable agreement with the calculation in Fig. 7e for $R_{eff} = 1.38$ m, whereas the straight fiber calculation in Fig. 7f, with $R_{eff} = \infty$, differs substantially from the experiment. The remaining discrepancy between the lowest experimental and calculated axial eigenfrequencies, highlighted by the blue circle in Fig. 7e, reflects a flattening of the CF profile near the contact region due to fiber deformation under contact pressure, including that caused by electrostatic and van der Waals forces [30, 34, 35]. In Section S6 of SI we investigate the related geometric and photoelastic effects of fiber squeezing. Our estimates show that squeezing by 1–10 nm can shift the CF by approximately 0.1–1 GHz comparable to the observed discrepancy.

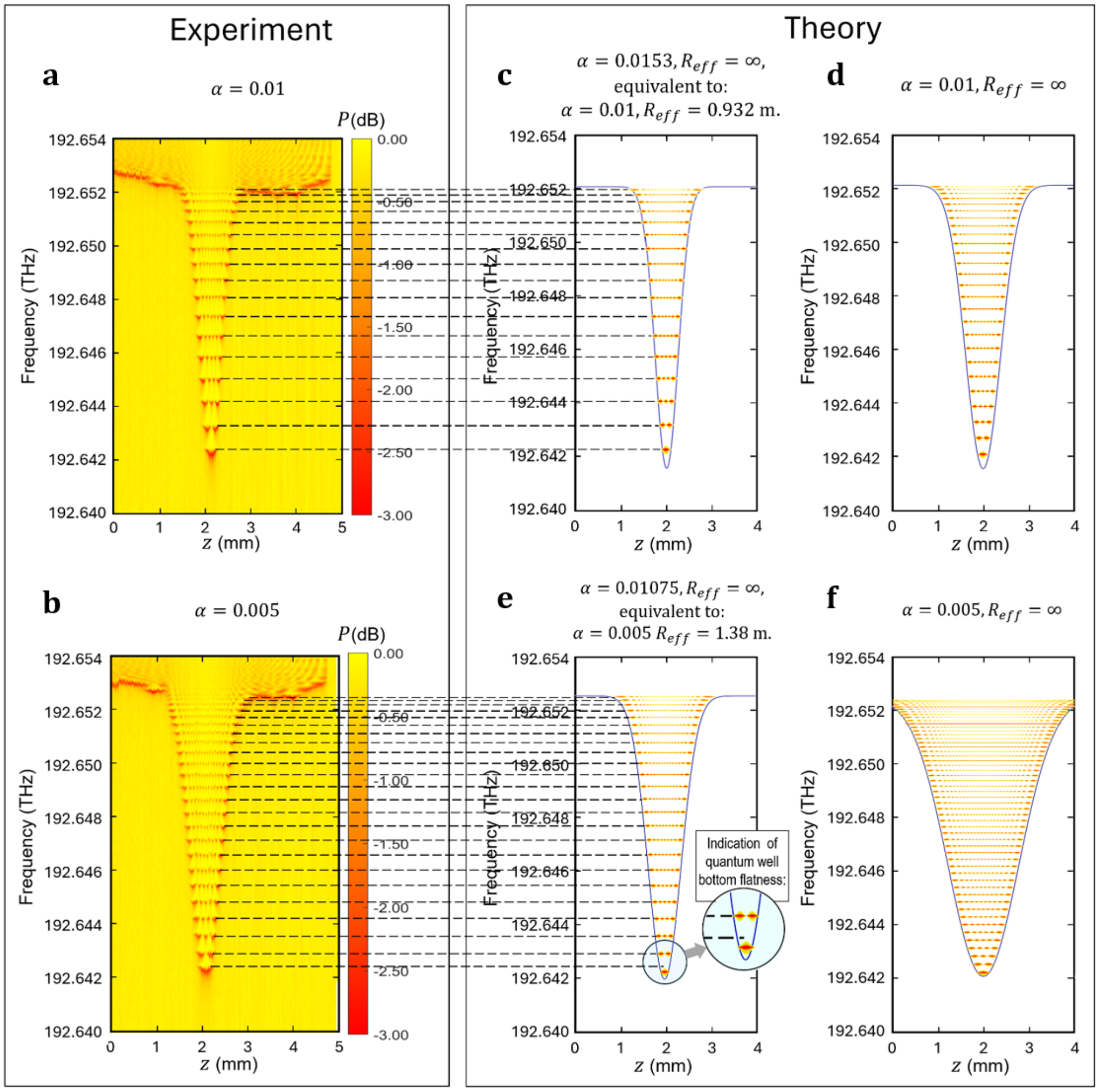

**Fig. 7.** Comparison of the experimental spectrograms from Ref. [30], where the vertical wavelength coordinates are rescaled to the frequency coordinates, with theoretical CF profiles and spectra for the same inter-fiber angles $\alpha$. a, b Experimental spectrograms for $\alpha = 0.01$ and $\alpha = 0.005$, respectively. c, d, e, f The corresponding theoretical CF profiles and spectra for c $\alpha = 0.01$ and $R_{eff} = 0.932$ m, d $\alpha = 0.01$ and $R_{eff} = \infty$, e $\alpha = 0.005$ and $R_{eff} = 1.38$ m, and f $\alpha = 0.005$ and $R_{eff} = \infty$.

## DISCUSSION

We theoretically described optical microresonators formed at the intersection of optical fibers and analyzed the physical effects underlying their shape and spectral characteristics. The CF variation of the induced microresonators is of the same order of magnitude as that in SNAP microresonators, which typically exhibit nanoscale effective radius variations along the optical fiber axis [17]. This allows us to attribute the described microresonators to the class of SNAP microresonators.

Two complementary coupling regimes have been considered. In the case of resonant interaction between closely spaced CFs, the intersecting fiber system can be described by a two-state model. In this approximation, coupling between two adjacent CFs of the fibers leads to the well-known effect of level repulsion [36], resulting in the formation of CF potential barriers and potential wells. At small fiber separations, we used numerical calculations to correct the analytical CF profile described by Eqs. (7) and (8). Since segments with sufficiently close CFs can be selected from the same parent fiber [37], the pure two-state coupling regime could be explored experimentally at separations where contributions from other modes remain negligible. In contrast, in the most common case when the induced CF shifts are small compared to the separation of CFs, their variation can be described perturbatively, leading to the sum of small corrections accumulated from many interacting states. Then, the CF variation has a universal Gaussian shape described by Eq. (14). Similar to the two-state model, the latter approximation fails at small fiber separations when the inter-fiber coupling becomes comparable to the CF separation. For the examples considered, the dominant contributions come from a sequence of WGMs in which increasing radial order is accompanied by decreasing azimuthal order, as illustrated in Fig. S5 of the SI. In both cases, a CF potential well localizes WGMs along the axial direction and forms a SNAP microresonator.

The comparison between the developed theory and experimentally measured spectrograms of microresonators induced at the intersection of optical fibers [30] showed that inclusion of the dramatically small but unavoidable curvature of the fiber segments is essential for achieving quantitative agreement with the measured spectra. While models neglecting curvature reproduce only the qualitative features of coupling-induced localization, they fail to account for the observed CF profiles and eigenfrequency distributions. By incorporating the meter-scale effective fiber curvature radius into the geometrical description of the fibers and the resulting coupling coefficients, the developed theory allows to explain the experimentally observed CF frequency variation and spectral characteristics of the induced microresonators.

It is also interesting to investigate losses of microresonator eigenstates induced by coupling to other fiber modes, as well as nonadiabatic effects at larger tilt angles or smaller curvature radii, where the coupling coefficients vary more rapidly along the axial coordinate $z$. Future studies may also address more complex configurations involving the intersection of fibers with pre-inscribed SNAP microresonator structures, as well as multiple intersecting fibers, thereby opening new opportunities for realizing SNAP devices with enhanced tunability and reconfigurability.

**Data availability**

The data that support the plots of this paper and other findings within this study are available from the author upon reasonable request.

**Funding**

This work was supported by the Engineering and Physical Sciences Research Council (EPSRC) (grants EP/X03772X/1 and EP/W002868/1).

**Competing interests**
The author declares no competing interests.

**Additional Information**
The details of calculations are presented in the Supporting Information.
**Correspondence** and requests for materials should be addressed to M. Sumetsky

**SUPPORTING INFORMATION**

# Microresonators induced at the optical fiber intersections

M. Sumetsky
*Aston Institute of Photonic Technologies, Aston University, Birmingham B4 7ET, UK*
m.sumetsky@aston.ac.uk

**S1. WGMS in a single fiber**

Here we consider WGMs of uncoupled fiber. Separating variables in local cylindrical coordinates $(\rho,\theta,z)$, we define the solution of the wave equation as $E_{m,q}(\mathbf{r}) = \exp(i\beta(\omega)z)\exp(im\theta)Q_q(\rho)$, where $Q_q(\rho)$ satisfies the equation [S1]

$$\frac{d^2Q}{d\rho^2}+\frac{1}{\rho}\frac{dQ}{d\rho}+\left(\frac{\omega^2}{c^2}n^2(\rho)-\frac{m^2}{\rho^2}\right)Q=0\,, \tag{S1}$$

corresponding to cutoff frequencies, for which $\beta(\omega) = 0$. For WGMs of our concern, we are only interested in fundamental and close to fundamental eigenstates of this equation having quantum numbers $q = 1,2, \ldots$. These states are localized near the fiber surface, $\rho = r_0$, where Eq. (S1) is simplified to the equation:

$$\frac{d^2Q}{dx^2}+\frac{1}{r_0}\frac{dQ}{dx}+\left(\frac{\omega^2}{c^2}n^2(x)-\frac{m^2}{r_0^2}+\frac{2m^2}{r_0^3}x\right)Q=0\,, \tag{S2}$$

where $x = \rho - r_0$ and

$$n(x)=\begin{cases} n_0, & x<0 \\ 1, & x>0 \end{cases} \tag{S3}$$

Solution of this equation in the region $\rho < r_0$ is expressed through the Airy function:

$$Q_q(\rho)=A_q\cdot \mathrm{Ai}\left(\varsigma(r_0-\rho)+a_q\right),\quad \varsigma=\frac{(2m^2)^{1/3}}{r_0}\cong\left(\frac{2n_0^2\omega^2}{c^2r_0}\right)^{1/3}, \tag{S4}$$

$$a_1=-2.338,\quad a_2=-4.088,\quad a_3=-5.521,\quad a_q \underset{q\gg1}{\simeq} -\left(\frac{3\pi}{2}\left(q-\frac{1}{4}\right)\right)^{2/3}, \tag{S5}$$

where $a_q$ are the roots of Airy function and we assumed that $Q_q(r_0) = 0$ , which is known to be in good agreement with numerical modelling [S2]. Remarkably, the asymptotic expression for the roots $a_q$ in Eq. (S5) is accurate with better than 1% starting from $q = 1$. Using this asymptotic, the normalization of the solution in the fiber cross-section $U(\rho,\theta) = \exp(im\theta)Q_q(\rho)$, which is reduced to the condition $2\pi r_0 \int_{-\infty}^{r_0} d\rho Q_q^2(\rho) = 1$, yields:

$$A_q=\left(\frac{\varsigma}{2r_0}\right)^{1/2}\left[\frac{3\pi}{2}\left(q-\frac{1}{4}\right)\right]^{-1/6}. \tag{S6}$$

Similar to the asymptotic for $a_q$ from Eq. (S5), this asymptotics is accurate with better than 1% starting with $q = 1$.

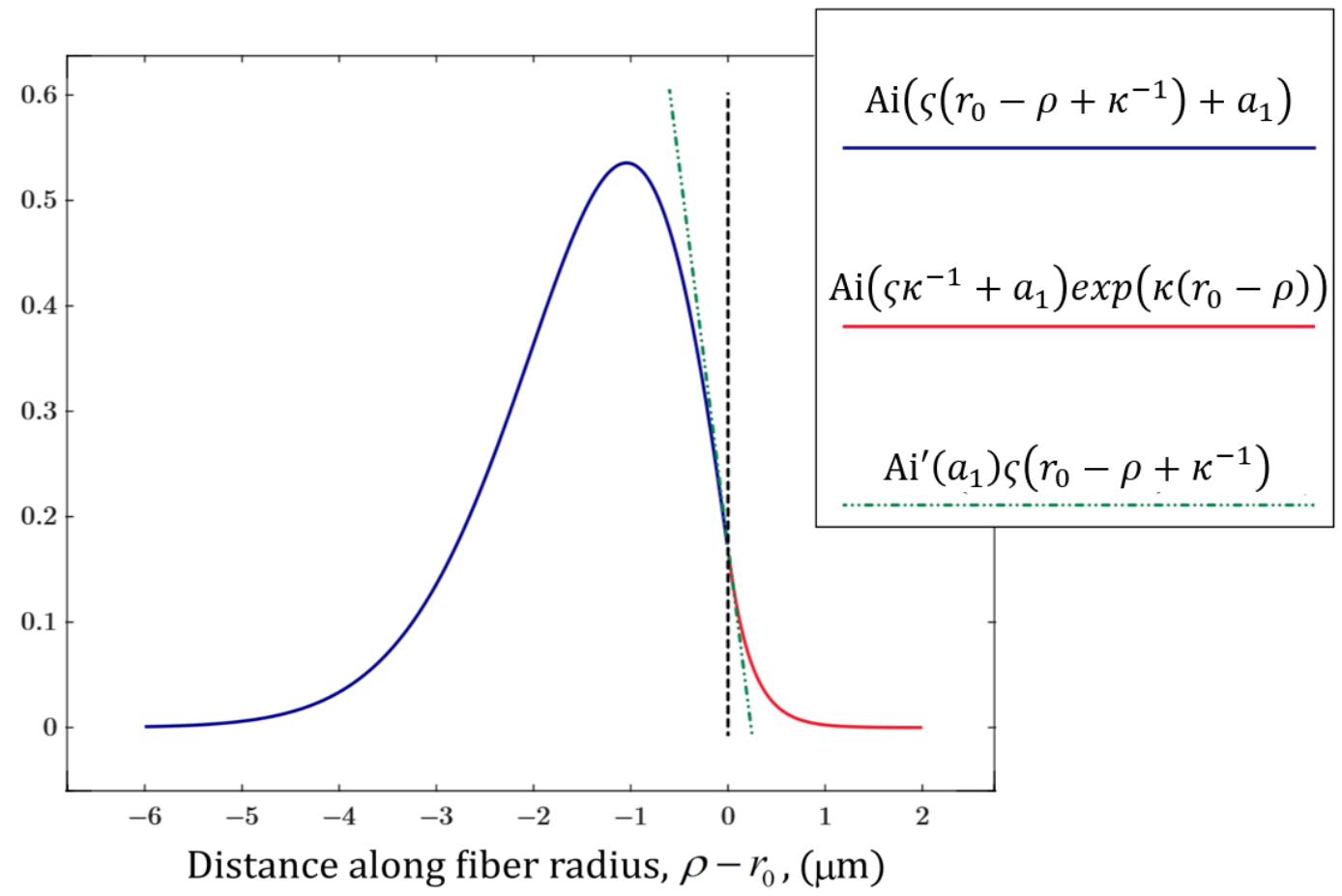

**Fig. S1.** Distribution of the fundamental WGM along the fiber radial direction. The distribution inside the fiber, expressed through the Airy function (blue), is matched with the exponential distribution outside the fiber (red).

Next, we abandon the assumption $Q_q(r_0) = 0$ and match the solution of Eq. (S1) inside the fiber with the evanescent solution at $\rho > r_0$ (Fig. S1). In this case, the solution near the fiber surface, $\rho = r_0$, is slightly shifted from zero and can be presented in the form:

$$Q_q(\rho) = A_q \cdot \mathrm{Ai}\left(\varsigma(r_0-\rho)+a_q+\Delta a_q\right) \tag{S7}$$

with the same normalization constant $A_q$ defined by Eq. (S6). Assuming that the solution considered is a TE mode, we match this solution and its derivative at $\rho = r_0$ with the evanescent solution at $\rho > r_0$,

$$Q_q(\rho) = B_q \exp\left(-\kappa(\rho-r_0)\right), \quad \kappa = \left(\frac{m^2}{r_0^2}-\frac{\omega^2}{c^2}\right)^{1/2} = \frac{\omega}{c}\left(n_0^2-1\right)^{1/2}, \tag{S8}$$

leads to the equation for $\Delta a_q$ and expression for $B_q$:

$$\frac{\mathrm{Ai}'\left(a_q+\Delta a_q\right)}{\mathrm{Ai}\left(a_q+\Delta a_q\right)} = \frac{\kappa}{\varsigma}, \quad B_q = A_q \mathrm{Ai}\left(a_q+\Delta a_q\right). \tag{S9}$$

Assuming $\Delta a_q \ll 1$, we find:

$$\begin{aligned} &\mathrm{Ai}(a_q+\Delta a_q) \cong \mathrm{Ai}'(a_q)\Delta a_q, \\ &\mathrm{Ai}'(a_q) \cong (-1)^{q-1}\pi^{-1/2}\left[\frac{3\pi}{2}\left(q-\frac{1}{4}\right)\right]^{1/6} \end{aligned} \tag{S10}$$

and Eq. (S9) is simplified:

$$\Delta a_q = \frac{\varsigma}{\kappa}, \quad B_q = (-1)^{q-1}\left(\frac{\varsigma^3}{2\pi r_0}\right)^{1/2}\frac{1}{\kappa}. \tag{S11}$$

From Eqs. (S4) and (S8)-(S11), we find for the evanescent field at $\rho > r_0$:

$$Q_q(\rho) = \frac{(-1)^{q-1} n_0}{\pi^{1/2}\left(n_0^2-1\right)^{1/2} r_0}\exp\left(-\kappa(\rho-r_0)\right). \tag{S12}$$

Remarkably, the magnitude of this field is *independent of the radial quantum number* $q$.

From Eqs. (S4), (S6), and (S10)-(S12), we find inside the fiber near its surface:

$$\begin{aligned} Q_q(\rho) &= A_q \cdot \mathrm{Ai}\left(\varsigma(r_0-\rho)+a_q+\Delta a_q\right) \cong A_q \cdot \mathrm{Ai}'\left(a_q\right)\left(\varsigma(r_0-\rho)+\Delta a_q\right) \\ &\cong (-1)^{q-1}\pi^{-1/2}\frac{n_0\omega_0}{c r_0}\left((r_0-\rho)+\frac{1}{\kappa}\right) \end{aligned} \tag{S13}$$

## S2. Coupled wave equations

The WGMs $E(\mathbf{r})$, which propagate along the configuration of fibers illustrated in Fig. 2, satisfy the wave equation, Eq. (4), of the main text. For each of the fibers, it is convenient to use the cylindrical curvilinear coordinates $(\rho_j, \theta_j, s_j)$, where $s_j$ coincides with the axis of the fiber $j$. Then, for a small deformation assumed and close to the fiber contact point corresponding to $s_1 = s_2 = 0$, we have

$$\begin{aligned}
\rho_2 &= 2r_0 + \Delta h_x - \rho_1 + \frac{1}{2}\left(\frac{1}{R_2} - \frac{1}{R_1}\right)z^2 + \frac{\alpha^2 z^2}{2r_0} + r_0\theta_1^2 + \alpha\theta_1 z \\
\theta_2 &= \pi - \theta_1 + \frac{\alpha z}{r_0} \\
s_2 &= s_1 = z
\end{aligned} \tag{S14}$$

In the approximation considered, the coordinates along the fiber axes $s_j$ coincide with the coordinate along the straight axis $z$ (see Fig. 2). Assuming first that each of the fibers stands alone, we can write down the WGM solution for the fiber $j$ as $E_j^{(0)}(\mathbf{r}) = \exp(i\beta_j(\omega)z)U_j(\rho_j, \theta_j)$ where the transverse component $U_j(\rho_j, \theta_j) = \exp(im_j\theta_j)Q_{jq}(\rho_j)$ satisfies the 2D wave equation [S1]

$$\Delta_{T_j} U_j(\rho_j, \theta_j) + \left(\frac{\omega^2}{c^2} n_j^2(\rho_j, \theta_j) - \beta_j^2(\omega)\right) U_j(\rho_j, \theta_j) = 0 . \tag{S15}$$

Here $n_j(\rho_j, \theta_j)$ is the refractive index distribution of the fiber $j$ standing alone, and the dependence of $\beta_j(\omega)$ on the angular frequency of light $\omega$ is determined from the condition of vanishing of $Q_{jq}(\rho_j)$ at large $\rho_j$. For the assumed close proximity to the CFs $\omega_{c,1}$ and $\omega_{c,2}$, we have [S3]

$$\beta_j^2(\omega) = 2\omega_0 \frac{n_0^2}{c^2}\left(\omega - \omega_{c,j}\right) . \tag{S16}$$

Here $\omega_0 \cong \omega_{cj}$ is the frequency in the narrow bandwidth considered.

We assume that the WGM CFs of fiber 1 and fiber 2, $\omega_{c,1}$ and $\omega_{c,2}$, are close to each other and to the frequency of light $\omega$ and separated from other CFs, so that we can ignore the effect of coupling with other transverse states. Then, the solution of the wave equation, Eq. (4), can be found in the form

$$E(\mathbf{r}) = \Psi_1(z)U_1(\rho_1, \theta_1) + \Psi_2(z)U_2(\rho_2, \theta_2) . \tag{S17}$$

In the calculations below, we assume that $U_j(\rho_j, \theta_j) = \exp(im\theta_j)Q_{jq}(\rho_j)$ where the radial component $Q_{jq}(\rho_j)$ is defined in Section S1.

Substituting Eq. (S17) into Eq. (4) and using Eq. (S15) for $j = 1$ and 2, we find:

$$\begin{aligned}
&\left[\beta_1^2(\omega) + k_0^2\left(n^2(\mathbf{r}) - n_1^2(\rho_1, \theta_1)\right)\right]U_1(\rho_1, \theta_1)\Psi_1(z) + \\
&\left[\beta_2^2(\omega) + k_0^2\left(n^2(\mathbf{r}) - n_2^2(\rho_2, \theta_2)\right)\right]U_2(\rho_2, \theta_2)\Psi_2(z) + \\
&U_1(\rho_1, \theta_1)\frac{d^2\Psi_1(z)}{dz^2} + U_2(\rho_2, \theta_2)\frac{d^2\Psi_2(z)}{dz^2} = 0
\end{aligned} \tag{S18}$$

Next, using Eq. (S14) we express $U_2$ in Eq. (S18) through coordinates $(\rho_1, \theta_1, z)$ as $U_2(\rho_2(\rho_1, \theta_1 z), \theta_2(\rho_1, \theta_1, z))$. Multiplying Eq. (S18) by $U_1^*(\rho_1, \theta_1)$ and integrating over the plane $(\rho_1, \theta_1)$ and, similarly, multiply it by $U_2^*(\rho_2, \theta_2) = U_2^*(\rho_2(\rho_1, \theta_1, z), \theta_2(\rho_1, \theta_1, z))$ and integrating over the plane $(\rho_1, \theta_1)$, we arrive at the coupled wave equations:

$$\begin{aligned}
\frac{d^2\Psi_1}{dz^2} + \beta_1^2(\omega)\Psi_1 &= -\frac{\omega_0^2}{c^2} I_{12}(z)\Psi_2 \\
\frac{d^2\Psi_2}{dz^2} + \beta_2^2(\omega)\Psi_2 &= -\frac{\omega_0^2}{c^2} I_{21}(z)\Psi_1
\end{aligned} \tag{S19}$$

In these equations, for the assumed small bending and tilting of fibers (see Eqs. (S14)), we introduced the dimensionless coupling parameters:

$$I_{12}(z) = I_{21}^* = \left(n_0^2 - 1\right)\int_0^{r_0}\int_{\theta_1} Q_1(\rho_1, \theta_1)Q_2^*(\rho_2(\rho_1, \theta_1, z), \theta_2(\rho_1, \theta_1, z))\rho_1 d\theta_1 d\rho_1 \tag{S20}$$

where $Q_j(q_j, \theta_j, z)$ are the normalized solutions defined in Section S1. Using Eq. (S15) for $j = 1,2$ and for the case $\beta_j(\omega) \ll n_j\omega/c$ of our interest (see Eq. (S16)), the integral over the cross-section of fiber 1 in Eq. (S20) can be transformed to the integral along the axis $\ell_\perp$ normal to the axis $v$ connecting the fiber centers as shown in Fig. S2:

$$I_{12}(z) = I_{21}^*(z) = \frac{c^2}{\omega^2}\int_{\ell_\perp}\left(Q_1^*\frac{\partial Q_2}{\partial v} - Q_2^*\frac{\partial Q_1}{\partial v}\right)d\ell_\perp\,. \tag{S21}$$

To calculate this integral, we assume that the WGMs in fibers have the same azimuthal quantum numbers $m_j$ and radial quantum numbers $q_j$ choose the axis $\ell_\perp$ at equal distances from the fibers where $Q_2 = Q_1$ and $\partial Q_2/\partial v = -\partial Q_1/\partial v$. Then, using Eq. (S12) for the normalized solution $Q_j$ we arrive at Eq. (7) of the main text.

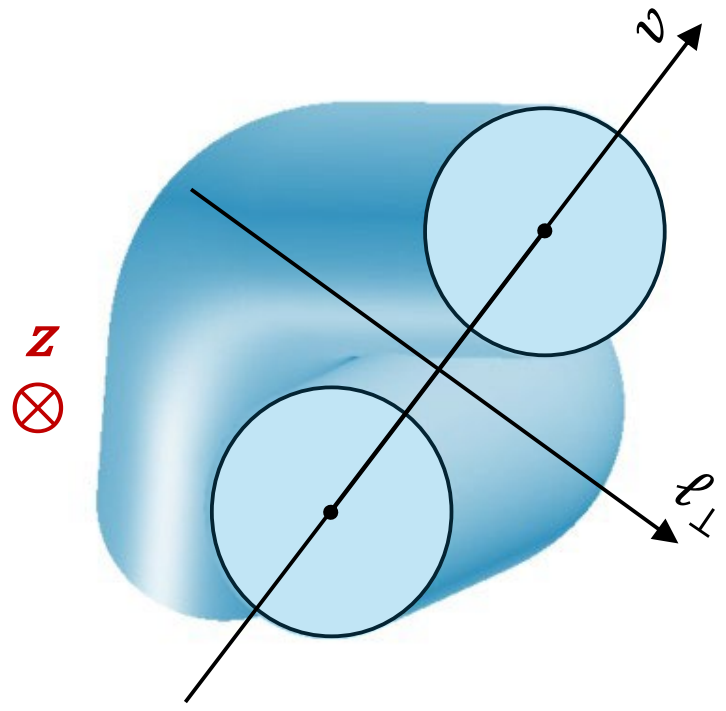

**Fig. S2**. Integration axis $l_\perp$ in Eq. (S21).

Coupled wave equations, Eqs. (S19), can be reduced to the fourth-order equation for $\Psi_1(z)$:

$$\frac{d^4\Psi_1}{dz^4} + a_3(z)\frac{d^3\Psi_1}{dz^3} + a_2(z)\frac{d^2\Psi_1}{dz^2} + a_1(z)\frac{d\Psi_1}{dz} + a_0(z)\Psi_1 = 0\,, \tag{S22}$$

where

$$\begin{aligned}
a_0(z,\omega) &= \beta_1^2(\omega)\left[\beta_2^2(\omega) + 2\left(\frac{1}{I_{12}(z)}\frac{dI_{12}(z)}{dz}\right)^2 - \frac{1}{I_{12}(z)}\frac{d^2I_{12}(z)}{dz^2}\right] - \frac{\omega_0^4}{c^4}I_{12}(z)I_{21}(z)\\
a_1(z,\omega) &= -2\frac{\beta_1^2(\omega)}{I_{12}(z)}\frac{dI_{12}(z)}{dz}\\
a_2(z,\omega) &= \beta_1^2(\omega) + \beta_2^2(\omega) + 2\left(\frac{1}{I_{12}(z)}\frac{dI_{12}(z)}{dz}\right)^2 - \frac{1}{I_{12}(z)}\frac{d^2I_{12}(z)}{dz^2}\\
a_3(z,\omega) &= -\frac{2}{I_{12}(z)}\frac{dI_{12}(z)}{dz}
\end{aligned} \tag{S23}$$

## S3. WKB and adiabatic solution of coupled wave equations

Assuming that the coefficients in Eqs. (S22) are sufficiently slow functions of $z$, we look for the solution of this equation in the WKB approximation. For this purpose, we formally add the factors $\varepsilon^n$ in front of $n^{\text{th}}$ derivatives in Eqs. (S22) and (S23) and look for the solution of Eq. (S22) in the form

$$\Psi_1(z) = \left(A_0(z) + \varepsilon A_1(z) + \ldots\right)\exp\left(\frac{i}{\varepsilon}\int^z \beta(z,\omega)dz\right) \tag{S24}$$

In the zero and first orders in $\varepsilon$, we arrive at the quartic algebraic equation for the propagation constant $\beta(z,\omega)$,

$$\beta^4(z,\omega) - \left(\beta_1^2(\omega) + \beta_2^2(\omega)\right)\beta^2(z,\omega) + \beta_1^2(\omega)\beta_2^2(\omega) - k_0^2 I_{12}(z)I_{21}(z) = 0 \tag{S25}$$

and the first-order linear differential equation for the preexponent $A_0(z,\omega)$,

$$\frac{dA_0}{dz} + P(z,\omega)A_0 = 0, \tag{S26}$$

$$P(z,\omega)=\frac{\left(a_{20}(z,\omega)-6\beta^2(z,\omega)\right)\dfrac{d\beta(z,\omega)}{dz}+a_1(z,\omega)\beta(z,\omega)-a_3(z,\omega)\beta^3(z,\omega)}{2\beta(z,\omega)\left(a_{20}(z,\omega)-2\beta^2(z,\omega)\right)}. \tag{S27}$$

where

$$\begin{aligned} a_{00}(z,\omega) &= \beta_1^2(\omega)\beta_2^2(\omega)-k_0^4 I_{12}(z)I_{21}(z)\\ a_1(z,\omega) &= 2\beta_1^2(\omega)\left[\mu(\omega)z-i\beta_0\right]\\ a_{20}(z,\omega) &= \beta_1^2(\omega)+\beta_2^2(\omega)\\ a_3(z,\omega) &= \mu(\omega)z-2i\beta_0\\ \mu(\omega) &= \frac{\omega(n_0^2-1)^{1/2}}{c}\left(\frac{1}{R_{eff}}+\frac{\alpha^2}{2r_0}\right) \end{aligned} \tag{S28}$$

The propagation constants of the coupled fiber system $\beta\left(z,\omega\right)=\beta_{\pm}\left(z,\omega\right)$ are found from Eq. (S25) as:

$$\beta_{\pm}^2(z,\omega)=\frac{1}{2}\left(\beta_1^2(\omega)+\beta_2^2(\omega)\right)\pm\sqrt{\frac{1}{4}\left(\beta_1^2(\omega)-\beta_2^2(\omega)\right)^2+\frac{\omega_0^4}{c^4}I_{12}(z)I_{21}(z)} \tag{S29}$$

Substituting the dependencies of $\beta_j(\omega)$ from Eq. (S16), we rewrite Eq. (S29) as

$$\beta_{\pm}^2(z,\omega)=\omega_0\frac{n_0^2}{c^2}\left(\left(2\omega-\omega_{c1}-\omega_{c2}\right)\pm\sqrt{\left(\omega_{c1}-\omega_{c2}\right)^2+\frac{\omega_0^2}{n_0^4}I_{12}(z)I_{21}(z)}\right) \tag{S30}$$

The CFs $\omega_c^{\pm}(z)$ of the system considered given by Eq. (8) of the main text can be simply found from this equation by setting $\beta_{\pm}^2(z)=0$.

Since the coefficient $P(z,\omega)$ in Eq. (S26) for the pre-exponent $A_0(z)$ is invariant under the change of the sign of the propagation constant $\beta(z,\omega)$, pre-exponent $A_0(z)$ is invariant under this change as well. Consequently, the eigenmodes of a SNAP resonator confined by the induced CF variations $\omega_c^{\pm}(z)$ can be constructed as a linear combination of the WKB solutions $A_0(z)\sum_{\pm}exp\left(\pm i\int^z\beta(z,\omega)dz\right)$ and, in particular, presented in the form

$$A_0(z)\cos\left(\int_{z_1}^{z}\beta_{\pm}(z,\omega)ds+\varphi_0\right). \tag{S31}$$

where the phase $\varphi_0$ is determined by matching this solution with exponentially decaying solutions away from the turning points (zeros of the propagation constant $\beta(z,\omega)$) $z_1$ and $z_2$ [S4].

In the adiabatic approximation, we ignore all derivatives in Eqs. (S23), i.e., set

$$\begin{aligned} a_0(z,\omega) &= \beta_1^2(\omega)\beta_2^2(\omega)-\frac{\omega_0^4}{c^4}I_{12}(z)I_{21}(z)\\ a_1(z,\omega) &= 0\\ a_2(z,\omega) &= \beta_1^2(\omega)+\beta_2^2(\omega)\\ a_3(z,\omega) &= 0 \end{aligned} \tag{S32}$$

and simplify the expression for $P(z,\omega)$ in Eq. (S27) to

$$P(z,\omega)=\frac{\left(\beta_1^2(\omega)+\beta_2^2(\omega)-6\beta^2(z,\omega)\right)\dfrac{d\beta(z,\omega)}{dz}}{2\beta(z,\omega)\left(\beta_1^2(\omega)+\beta_2^2(\omega)-2\beta^2(z,\omega)\right)}. \tag{S33}$$

Then, the solution of Eq. (S26) is

$$A_0(z,\omega)=\frac{A_{00}}{\sqrt{\beta(z,\omega)\left(\beta_1^2(\omega)+\beta_2^2(\omega)-2\beta^2(z,\omega)\right)}}. \tag{S34}$$

where $\beta(z,\omega)=\beta_{\pm}(z,\omega)$ is determined by Eqs. (S30), or, equivalently, by Eqs. (8) and (9) of the main text.

## S4. Multipole eigenfrequency calculation method

In the adiabatic approximation, at a fixed surface gap $\Delta d(z) = \Delta d_0$, the cutoff frequencies of the coupled fibers are obtained as eigenfrequencies of the two-dimensional transverse problem (Fig. S3). Here we assume that the fiber radii $r_{01}, r_{02}$ and refractive indices $n_1, n_2$ may be different. The calculation considers each local cross-section as two circular, axially uniform dielectric cylinders. Contact deformation and photoelastic changes are considered separately in section S6.

For an isolated fiber $j = 1,2$, continuity of the axial TE electric field and its derivative along the fiber $j$ radial coordinate $r_j$ at $r_j = r_{0j}$ gives [S1]

$$n_j \frac{J_{m_j}'(n_j s_j)}{J_{m_j}(n_j s_j)} = \frac{H_{m_j}^{(1)'}(s_j)}{H_{m_j}^{(1)}(s_j)}, \quad s_j = k(\omega) r_{0j}, \quad k(\omega) = \frac{\omega}{c} \tag{S35}$$

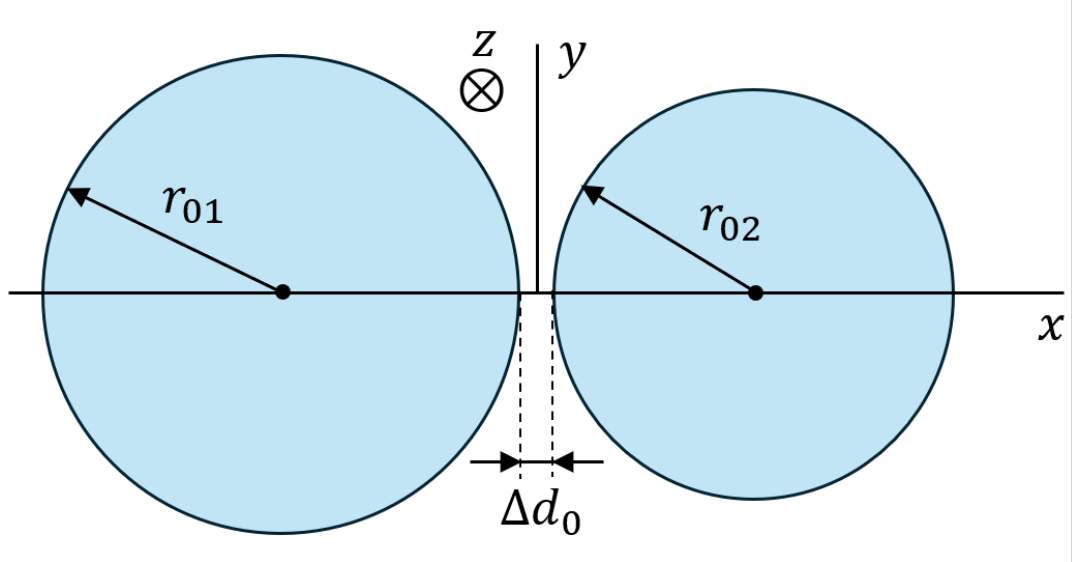

**Fig. S3**. Cross-section of fibers with radii $r_{01}$ and $r_{02}$ separated by a gap $\Delta d_0$.

Here $J_m$ and $H_m^{(1)}$ are the Bessel and outgoing Hankel functions, $m_j$ is the positive azimuthal order selected on fiber $j$, and the prime denotes the derivative of the entire argument.

For a system of two fibers, the multipole eigenfrequency calculation method (see e.g., [S5, S6]) used in our calculation is described as follows. We place two fibers side by side, with their axes separated by $d = r_{01} + r_{02} + \Delta d_0$. We expand the exterior field as outgoing cylindrical waves about both fiber axes and each interior field as regular Bessel waves about its own axis. Graf's addition theorem expresses the field outgoing from cylinder $k$ as regular harmonics incident on cylinder $j \neq k$ [S5-S7]. We define

$$\Lambda_{i,j\ell} = n_j \frac{J_\ell'(n_j s_j)}{J_\ell(n_j s_j)}, \quad \Lambda_{J,j\ell} = \frac{J_\ell'(s_j)}{J_\ell(s_j)}, \quad \Lambda_{H,j\ell} = \frac{H_\ell^{(1)'}(s_j)}{H_\ell^{(1)}(s_j)}, \quad D_{j\ell} = \frac{\Lambda_{H,j\ell} - \Lambda_{i,j\ell}}{\Lambda_{i,j\ell} - \Lambda_{J,j\ell}}, \tag{S36}$$

and

$$T_{jk,\ell\ell'}^{(h)} = \frac{J_\ell(s_j)}{H_\ell^{(1)}(s_j)} \left[ H_{\ell+\ell'}^{(1)}(k_0 d) + h(-1)^{\ell} H_{\ell-\ell'}^{(1)}(k_0 d) \right]. \tag{S37}$$

Here, the row number $\ell$ belongs to fiber $j$ and the column number $\ell'$ to fiber $k$, and $h = \pm 1$ defines the parity of solution along axis $y$. We consider the modes with quantum numbers within $\mathcal{L}_j = \{m_j - L, \dots, m_j + L\}$ on each fiber. Then, introducing $\mathbf{D}_j = \mathrm{diag}\{D_{j\ell}: \ell \in \mathcal{L}_j\}$ and the blocks $\mathbf{T}_{jk}^{(h)}$, we arrive at the eigenfrequency equation:

$$\det \begin{pmatrix} \mathbf{D}_1 & -\mathbf{T}_{12}^{(h)} \\ -\mathbf{T}_{21}^{(h)} & \mathbf{D}_2 \end{pmatrix} = 0. \tag{S38}$$

We solve this eigenfrequency equation numerically for a range of fibre gaps $\Delta d_0$. Starting at a sufficiently large gap, where the solutions approach the isolated-fibre eigenfrequencies defined by Eq. (S35), we track the roots as the gap is reduced.

Standard silica optical fibres have a dense spectrum of WGM eigenfrequencies in the frequency range of our interest. Figure S4 shows the isolated-fiber eigenfrequencies between 192 and 193.5 THz, including the mode with $m = 353$ and $q = 1$ used in our numerical comparisons. The refractive index of both fibers is $n_0 = 1.44$; the radii are $r_{01} = 62.5$ µm and $r_{02} = r_{01} + \Delta r$, with $\Delta r = 0,10,20,30$ nm. Even these small radius differences produce appreciable shifts of the eigenfrequencies, which are defined by rescaling $\Delta\omega/\omega_0 = -\Delta r/r_{01}$.

Among the many modes of fibers 1 and 2 in the vicinity of the mode with $m = 353$ and $q = 1$ shown in Fig. S4, only a relatively small subset contributes appreciably to the coupling-induced eigenfrequency splitting for equal fibers and to the eigenfrequency shift for unequal fibers as the gap $\Delta d_0$ is varied. Figure S5 shows, on a logarithmic scale, the magnitudes $|\Delta\omega|/(2\pi)$ of the frequency shifts of the $(m_1, q_1) = (353,1)$ mode of fiber 1 induced by coupling to individual modes $(m_2, q_2)$ of fiber 2. The dominant contributions lie near the frequency-matching curve $\omega^{(2)}_{m_2,q_2} \simeq \omega^{(1)}_{353,1}$. For unequal fibers, the number of modes of fiber 2 that contribute appreciably decreases only slightly as $\Delta d_0$ increases. For equal fibers, however, the contribution of the matching mode, $m_2 = m_1 = 353$ and $q_2 = q_1 = 1$, exceeds those of the other individual modes by factors of approximately 80, 14, 7 and 4 at $\Delta d_0 = 1, 0.5, 0.3$ and 0.1 µm, respectively. The next section S5 discusses the two-mode approximation, which follows from the equations presented here by retaining only the coupling between modes with $m_2 = m_1$ and $q_2 = q_1$.

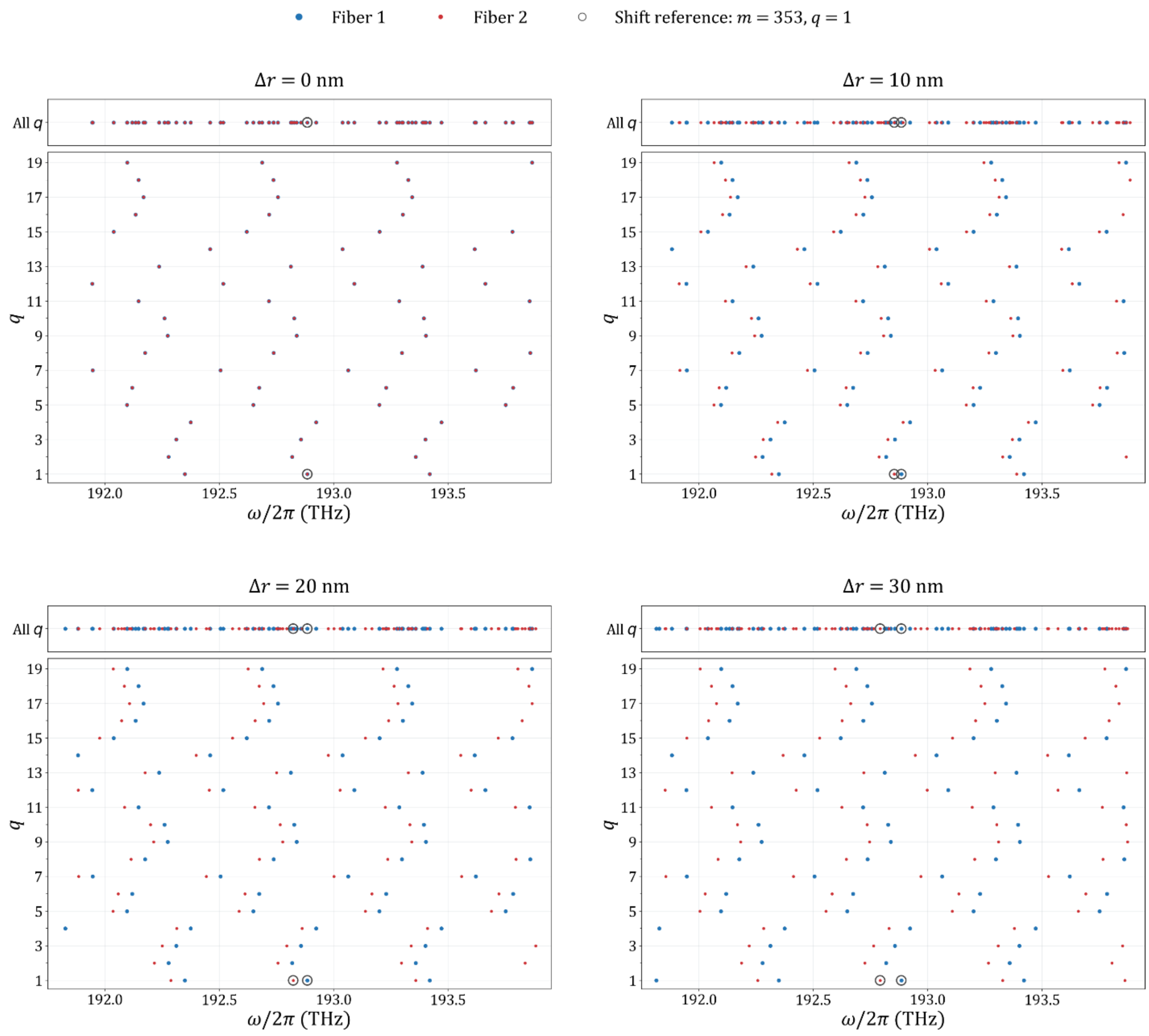

**Fig. S4.** The eigenfrequencies of isolated fibers in the bandwidth between 192 and 193.5 THz, including the mode with $m = 353$ and $q = 1$.

## S5. Cylindrical and planar weak coupling approximation, comparison with numerical calculations

We assume now that fibers 1 and 2 have close isolated eigenfrequencies, $\omega_{m_1}^{(1)}$ and $\omega_{m_2}^{(2)}$ with $q_2 = q_1 = 1$ and retain only one mode with quantum number $m_j$ in fiber $j$. Then Eq. (S38) is reduced to

$$D_{1m_1} D_{2m_2} - T_{12,m_1m_2}^{(h)} T_{21,m_2m_1}^{(h)} = 0 \tag{S39}$$

Using $H_{-q}^{(1)} = (-1)^q H_q^{(1)}$, setting $\omega_{m_1m_2}^{(c)} = \left(\omega_{m_1}^{(1)} + \omega_{m_2}^{(2)}\right)/2$, and linearizing $D_{jm_j}$ in frequency near $\omega_{m_j}^{(j)}$, we define the coupling $K_h$ in frequency units from the equation

$$K_h^2\left(\Delta d_0\right) = \frac{T_{12,m_1m_2}^{(h)}\left(\omega_{m_1m_2}^{(c)}, \Delta d_0\right) T_{21,m_2m_1}^{(h)}\left(\omega_{m_1m_2}^{(c)}, \Delta d_0\right)}{D_{1m_1}{}'\left(\omega_{m_1}^{(1)}\right) D_{2m_2}{}'\left(\omega_{m_2}^{(2)}\right)}, \tag{S40}$$

where

$$D_{jm_j}{}'\left(\omega_j\right) = \frac{dD_{jm_j}}{d\omega} = \frac{\pi s_{j0}^2\left(n_j^2 - 1\right) J_{m_j}\left(s_{j0}\right) H_{m_j}^{(1)}\left(s_{j0}\right)}{2i\omega_{m_j}^{(j)}}, \quad s_{j0} = \frac{\omega_j r_{0j}}{c}.$$

The roots of the reduced equation are

$$\omega_\pm\left(\Delta d_0\right) \simeq \frac{1}{2}\left(\omega_{m_1}^{(1)} + \omega_{m_2}^{(2)}\right) \pm \sqrt{\frac{\left(\omega_{m_1}^{(1)} - \omega_{m_2}^{(2)}\right)^2}{4} + K_h^2\left(\Delta d_0\right)}\,. \tag{S41}$$

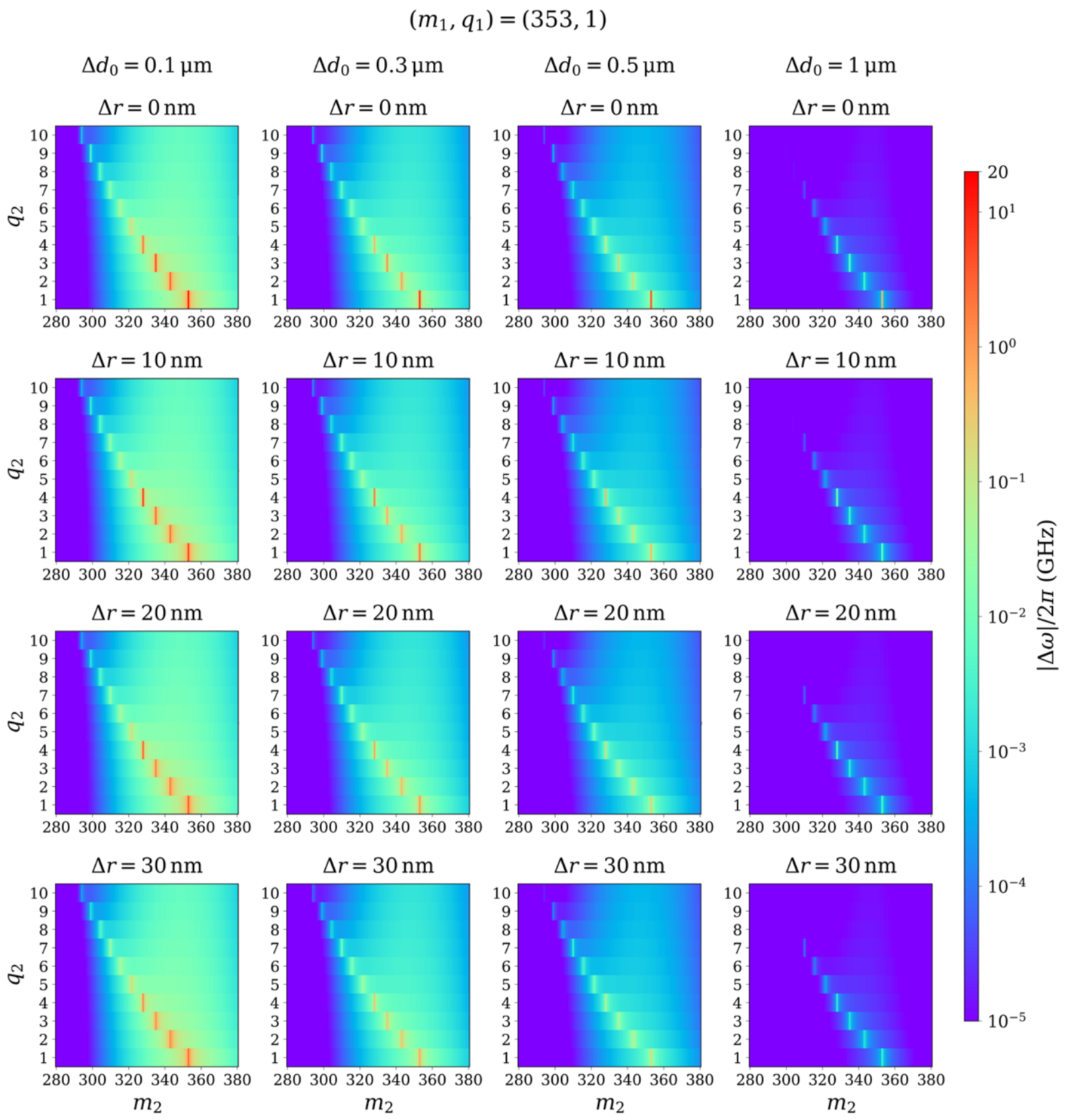

**Fig. S5.** the magnitudes $|\Delta\omega|/(2\pi)$ of the frequency shifts of the $(m_1, q_1) = (353,1)$ mode of fiber 1 induced by coupling to individual modes $(m_2, q_2)$ of fiber 2.

Neglecting the much smaller second term in brackets in the expression for $T^{(h)}_{ij,m_i m_j}$ in Eq. (S37), which is proportional to $h$, and the small difference between the isolated and mean frequencies yields

$$K_h\left(\Delta d_0\right) \simeq K\left(\Delta d_0\right) = \frac{2i\omega_j}{\pi s_1 s_2 \sqrt{\left(n_1^2-1\right)\left(n_2^2-1\right)}} \frac{H^{(1)}_{m_1+m_2}\left(k_0\left(r_{01}+r_{02}+\Delta d_0\right)\right)}{H^{(1)}_{m_1}\left(s_1\right) H^{(1)}_{m_2}\left(s_2\right)}, \quad s_j = k_0 r_{0j}. \tag{S42}$$

This expression for the coupling, expressed through the cylindrical functions, is called here the *cylindrical approximation*.

In the large-order approximation we have $H^{(1)}_N(Nt) \simeq -i\left[2/\left(\pi N\sqrt{1-t^2}\right)\right]^{1/2} e^{N\eta(t)}$, where $\eta(t) = \operatorname{arcosh}(1/t) - \sqrt{1-t^2}$ [S7] and Eq. (S42) can be simplified further. In particular, in the first order in $\Delta d_0/d_0$, we find

$$\left|K\left(\Delta d_0\right)\right| \simeq \frac{\omega_j}{s_1 s_2 \sqrt{\left(n_1^2-1\right)\left(n_2^2-1\right)}} \left[\frac{2m_1 m_2}{\pi\left(m_1+m_2\right)}\right]^{1/2} \left[\frac{\left(1-t_1^2\right)\left(1-t_2^2\right)}{1-t_0^2}\right]^{1/4} \exp\left(-\left(\left(\frac{m_1+m_2}{r_{01}+r_{02}}\right)^2 - k_0^2\right)^{1/2} \Delta d_0\right). \tag{S43}$$

where

$$t_j = \frac{s_j}{m_j} < 1, \quad t_0 = \frac{k_0\left(r_{01}+r_{02}\right)}{m_1+m_2} < 1. \tag{S44}$$

For almost equal fibers, we set $m_j = m \simeq n_0 k_0 r_0$, and the exponent in this expression is reduced to $\exp\left(-k_0\sqrt{n_0^2-1}\Delta d_0\right)$. The latter approximation, which we call here the *planar approximation*,

$$\left|K\left(\Delta d_0\right)\right| \simeq \frac{c^{3/2}}{\pi^{1/2} r_0^{3/2} \omega_0^{1/2} \left(n_0^2-1\right)^{3/4}} \exp\left(-\frac{\omega_0}{c}\sqrt{n_0^2-1}\Delta d_0\right). \tag{S45}$$

coincides, after rescaling by the factor $2n_0^2/\omega_0$, with the dimensionless coupling $I_{12}$ defined by Eq. (7) of the main text and considered in section S1. For comparison of analytical cylindrical (Eq. (S42)) and planar (Eq. (S45)) approximations with numerical calculations using the multipole method described in section S4, we set $r_{01} = 62.5\,\mu$m, $n_1 = n_2 = 1.44$, and $r_{02} = r_{01} + \Delta r$ with $\Delta r = 0,\ 10,\ 20,\ 30$ nm. At the first radial quantum number, $q_1 = 1$, we follow the pair of eigenfrequencies associated with $m_1 = m_2 = 353$ as $\Delta d_0$ changes from 2 µm to 0.01 µm. In our numerical calculations we consider truncations with $L = 20,\ 30,\ 40,\ 50$ (see section S4). Their comparison confirms the calculated eigenfrequency accuracy relative to the change of their separation, which is better than 2% for $\Delta d_0 > 0.1$ µm, and better than 1% for $\Delta d_0 > 0.2$ µm.

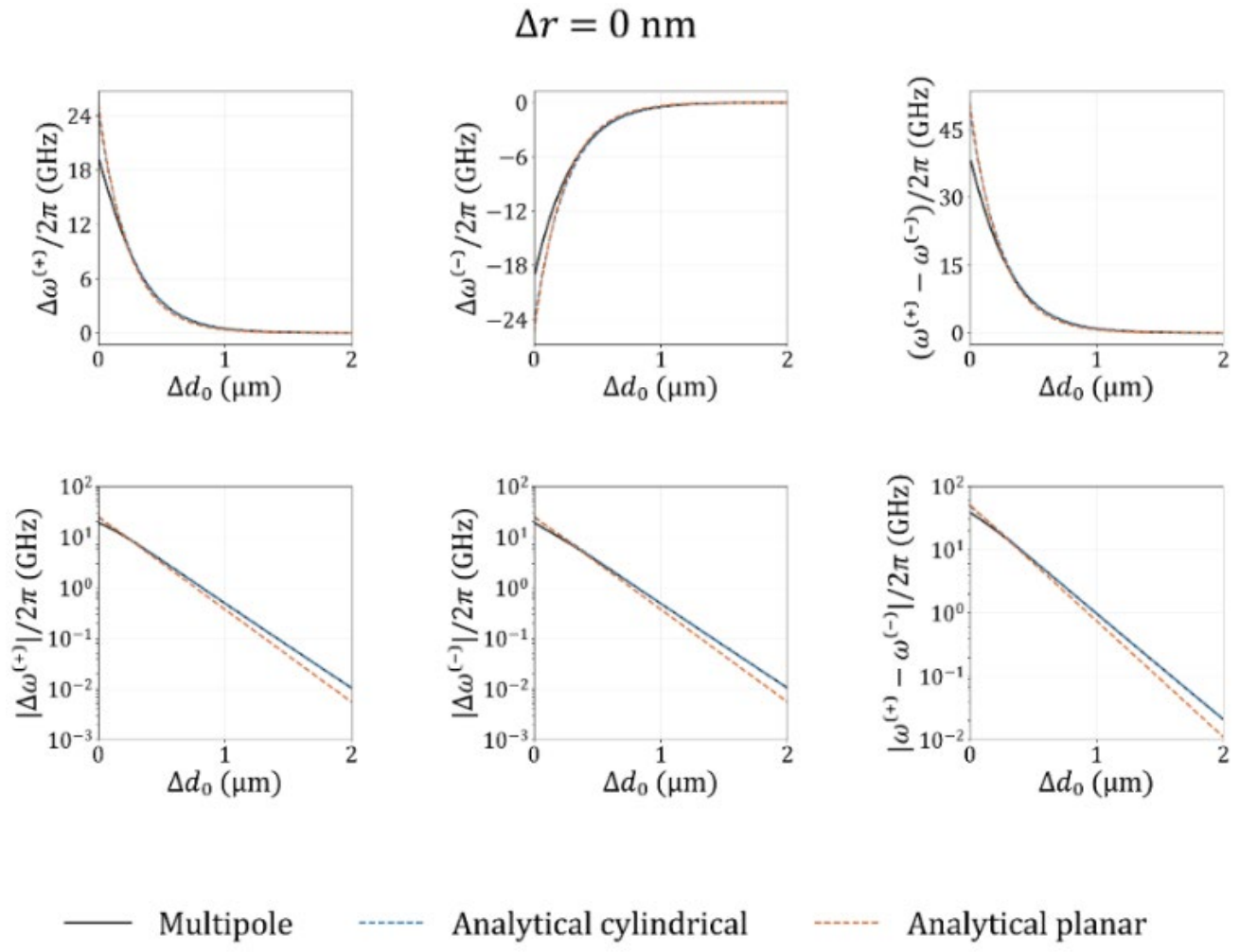

**Fig. S6**. Cutoff-frequency shifts $\Delta\omega^{(+)}/2\pi$ and $\Delta\omega^{(-)}/2\pi$ and the splitting $\left(\omega^{(+)} - \omega^{(-)}\right)/2\pi$ as functions of the fiber separation $\Delta d_0$ for identical fibers ($\Delta r = 0$): linear (upper row) and the corresponding logarithmic (lower row) scales.

Figure S6 shows the coupling-induced cutoff-frequency shifts $\Delta\omega^{(+)}/2\pi$ and $\Delta\omega^{(-)}/2\pi$ and the splitting $\left(\omega^{(+)} - \omega^{(-)}\right)/2\pi$ as functions of the fiber separation $\Delta d_0$ for identical fibers ($\Delta r = 0$) calculated by multipole method and compared with cylindrical and planar analytical approximations. The upper and lower rows show the results on linear and logarithmic scales, respectively. Absolute values are used in the logarithmic plots. Although the two analytical predictions are close at small gaps, both overestimate the numerical splitting. At $\Delta d_0 = 0.01$ μm, the numerical, cylindrical, and planar splittings are 37.92, 47.34, and 48.87 GHz, respectively, giving analytical overestimates of approximately 25% and 29%. The cylindrical approximation improves rapidly with increasing separation: over gaps of 1-2 μm, its splitting and individual shifts differ from the numerical results by less than 0.03% and 0.6%, respectively. The planar expression decays more rapidly and underestimates the splitting by approximately 24-47% over the same interval.

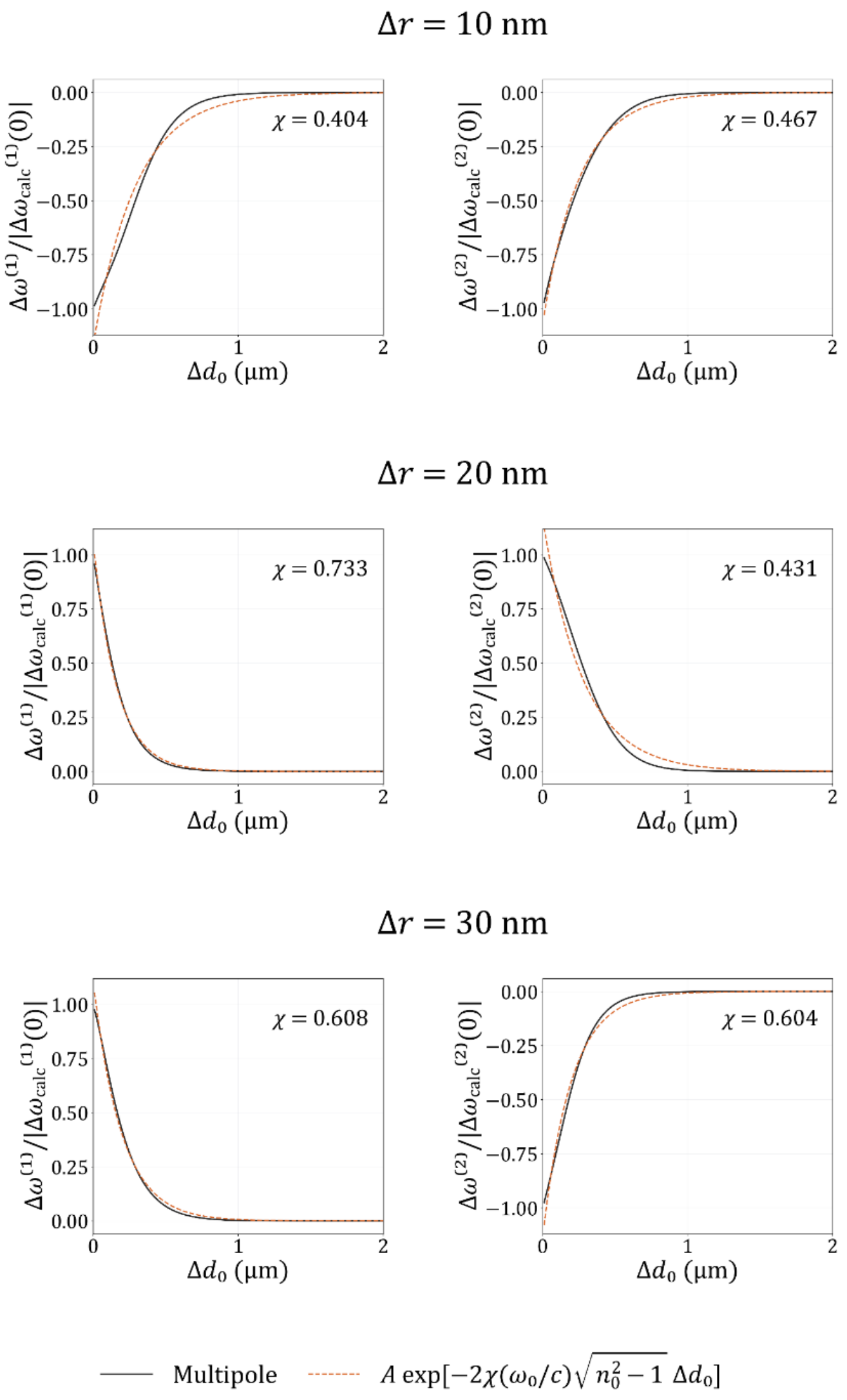

**Fig. S7.** Comparison of normalized cutoff-frequency shifts calculated by multipole method with fitted exponential dependencies for unequal fibers with radius differences $\Delta r = 10,\ 20,\ 30$ nm.

Figure S7 compares multipole calculations with fitted exponentials for the individual cutoff-frequency shifts of unequal fibers with radius differences $\Delta r = 10,\ 20,\ 30$ nm. The cylindrical and planar analytical approximations do not reproduce these shifts throughout the gap range considered. We therefore fit each branch independently to $A\exp\left(-2\chi(\omega_0/c)\sqrt{n_0{}^2-1}\Delta d_0\right)$. Here $\chi = 0.5$ corresponds to the eigenfrequency splitting in the planar approximation (Eq (S45)) when the eigenfrequencies of uncoupled modes are equal, while $\chi = 1$ corresponds to the case when the isolated eigenfrequencies are different and their splitting is much smaller than their separation in isolated fibers. Each curve is normalized by the magnitude of its numerical shift at the smallest calculated gap. The fits cover gaps of 0.01-2 μm and yield $\chi = 0.404 - 0.733$, quantifying differences in the effective gap dependence among the branches and radius mismatches. Across the six curves, the root-mean-square fit errors are 0.8-3.8% of the maximum shift magnitude, with maximum deviations of up to 15%. This comparison will be used for estimating the experimental parameters of microresonators induced by tilted fibers discussed in the main text.

## S6. Squeeze-induced cutoff frequency shift

Here we derive analytical estimates for two identical silica fibers squeezed symmetrically at their contact. We consider the 2D model, when the values of cutoff frequencies coincide with values of 2D eigenfrequencies.

We start with the purely *geometrical effects of local fiber deformation*. Let $\xi > 0$ be the total reduction of the center-to-center distance due to squeezing. In the flat-contact approximation, we assume that a circular cap of height $\xi/2$ is removed from each fiber and replaced by a straight contact segment [S8] (Fig. S8). For $\xi/r_0 \ll 1$, the contact half-width $a = \sqrt{r_0\,\xi}$. Using the WGM quantization condition $\omega n_0 L_f/c = 2\pi m$, where $m \gg 1$ is an integer and $L_f = 2\pi r_0$ is the fiber circumference, we determine the frequency shift due to the change of the circumference $\Delta L_f$ as $\delta\omega_{\text{geom}} = -\omega_0 \Delta L_f/L_f$.

For one fiber, the circular arc $2r_0\theta$ is replaced by the straight segment $2a = 2r_0\sin(\theta)$. For both fibers, the total circumference and its change are therefore $2L_f = 4\pi r_0$ and $2\Delta L_f = 2(2a - 2r_0\theta) = 4r_0(\sin(\theta) - \theta)$. Expanding $\sin(\theta)$, we find the frequency shift caused by geometric fiber deformation

$$\delta\omega_{\text{geom}} \simeq \frac{\omega_0}{6\pi}\left(\frac{\xi}{r_0}\right)^{3/2} \tag{S46}$$

The frequency shift determined by this equation is positive because replacing an arc by a chord (Fig. S8) shortens the optical path.

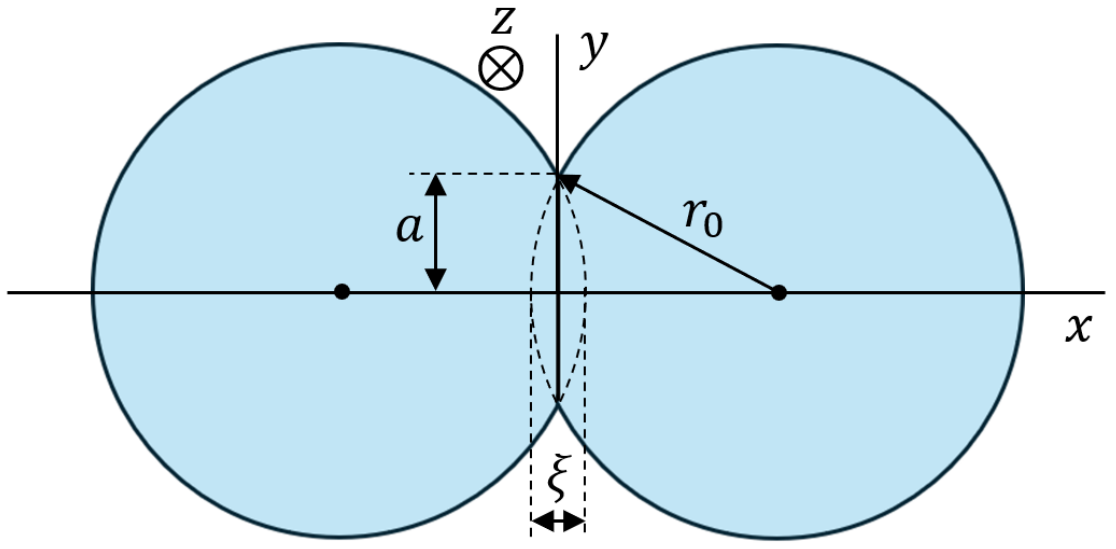

**Fig. S8**. Cross-section of identical fibers with radius $r_0$ squeezed by $\xi$.

Let us now consider the effects of the *photoelastic change in refractive index*. We choose axis $x$ normal to the contact surface, $y$ parallel to the contact surface, and $z$ along the fiber axis and along the electric field of the two-dimensional TE mode (Fig. S8). For isotropic silica, the index change for this polarization is $\Delta n_z = -\frac{1}{2}n_0^3\left[p_{11}S_{zz} + p_{12}\left(S_{xx} + S_{yy}\right)\right]$ where $S_{ij}$ are strain components. Here $p_{11}$ and $p_{12}$ are the strain-optic coefficients. In the simplest uniaxial estimate, $S_{xx} = S$ and the free lateral strains are $S_{yy} = S_{zz} = -\mu S$, where $\mu$ is the Poisson ratio. The average compressive strain is estimated as the indentation of one fiber divided by the contact depth $a$: $S \simeq -\xi/2a \simeq -\frac{1}{2}\sqrt{\xi/r_0}$. Consequently, we have

$$\Delta n_z \simeq \frac{n_0^3}{4}\sqrt{\frac{\xi}{r_0}}P_z, \quad P_z = p_{12}\left(1-\mu\right) - p_{11}\mu. \tag{S47}$$

For silica fibers, we have $p_{11} = 0.121$, $p_{12} = 0.270$, and $\mu = 0.17$ give $P_z = 0.2035$ [S9].

Let $E_j^{(0)}(\mathbf{r})$ be the electric field of the unperturbed exchange supermode $j = +$ or $j = -$, and let $\Omega_s$ denote the strained region inside the two fibers. For a refractive index perturbation $\Delta n_z(\mathbf{r})$ in silica fibers, first-order eigenfrequency perturbation theory gives for the photoelastic eigenfrequency shift

$$\delta\omega_j^{\mathrm{PE}} = -\frac{\omega_0 n_0}{U_j}\int_{\Omega_s}\Delta n_z(\mathbf{r})\left|E_j^{(0)}(\mathbf{r})\right|^2 dA, \quad U_j = \int n^2(\mathbf{r})\left|E_j^{(0)}\right|^2 dA. \tag{S48}$$

Here $U_j$ is the electric field normalization and $n(\mathbf{r}) = n_0$ inside fibers and $n(\mathbf{r}) = 1$ outside them. If the index change is replaced by its mean value $\overline{\Delta n_z}$ in $\Omega_s$, define the unperturbed modal-energy fraction

$$\delta\omega_j^{\mathrm{PE}} \simeq -\omega_0\frac{\overline{\Delta n_z}}{n_0}\eta_j, \quad \eta_j = \frac{n_0^2}{U_j}\int_{\Omega_s}\left|E_j^{(0)}\right|^2 dA. \tag{S49}$$

For symmetric weak coupling, the leading self-energy fractions of the two exchange parities are equal: $\eta_+ \simeq \eta_-$.

The fundamental radial WGM can be expressed through the Airy function as described in Section S1 which yields $\eta_j \simeq 2n_0^2\xi/[\pi r_0(n_0^2-1)]$ and

$$\delta\omega_{\mathrm{PE}} \simeq -\frac{n_0^4\omega_0\left[p_{12}(1-\mu)-p_{11}\mu\right]}{2\pi\left(n_0^2-1\right)}\left(\frac{\xi}{r_0}\right)^{3/2}. \tag{S50}$$

Adding the geometric and photoelastic contributions defined by Eqs. (S46) and (S50), we find the total frequency shift

$$\delta\omega \simeq \frac{\omega_0}{\pi}\left(\frac{\xi}{r_0}\right)^{3/2}\left\{\frac{1}{6}-\frac{n_0^4\left[p_{12}(1-\mu)-p_{11}\mu\right]}{2\left(n_0^2-1\right)}\right\}. \tag{S51}$$

Substituting the parameters for the silica fibers given above, we find within an order of magnitude

$$\delta\omega \simeq -0.1\omega_0\left(\frac{\xi}{r_0}\right)^{3/2}. \tag{S52}$$

Remarkably, the change of the exchange splitting is predominantly determined by the geometrical changes for finite $\xi$. To determine this change, we introduce the gap between unsqueezed fibers along coordinate $x$ near the contact point before squeezing, $\Delta h_0(y) = y^2/r_0$, and after squeezing (Fig. S8):

$$\Delta h_\xi(y) = \max\left(\frac{y^2}{r_0}-\xi, 0\right) \tag{S53}$$

Replacing the coordinate $z$ by the variable $\xi$ in Eq. (S21) for the overlap integral, $I_{12}(z) \to I_{12}(\xi)$, and introducing the frequency splitting $\Delta\omega(\xi)$, we find for $\kappa\xi \ll 1$

$$\frac{I_{12}(\xi)}{I_{12}(0)} = \frac{\Delta\omega(\xi)}{\Delta\omega(0)} = \left(\int_{-\infty}^{\infty} e^{-\kappa\Delta h_\xi(y)}dy\right)\Big/\left(\int_{-\infty}^{\infty} e^{-\kappa\Delta h_0(y)}dy\right) \simeq 1+\kappa\xi, \quad \kappa = \frac{\omega_0}{c}\left(n_0^2-1\right)^{1/2}. \tag{S54}$$

Using the expression for the asymptotic splitting found from Eqs. (S45)

$$\Delta\omega(0) \simeq \frac{2c^{3/2}}{\pi^{1/2} r_0^{3/2}\omega_0^{1/2}\left(n_0^2-1\right)^{3/4}}, \tag{S55}$$

we find

$$\delta\omega = \kappa\xi\Delta\omega(0) = \frac{2(c\omega_0)^{1/2}}{\pi^{1/2} r_0^{3/2}\left(n_0^2-1\right)^{1/4}}\xi. \tag{S56}$$

Using $r_0 = 62.5$ µm, $n_0 = 1.44$, and $\omega_0 = 2\pi \times 193$ THz, we obtain an evanescent decay rate $\kappa \cong 4.2$ µm$^{-1}$. Thus, Eq. (S54) predicts an increase in the inter-fiber coupling of approximately 0.4–4% for symmetric squeezing by $\xi$ = 1–10 nm. With an unsqueezed full CF splitting $\Delta\omega(0)/(2\pi) \simeq 50$ GHz, the corresponding increase in splitting is approximately 0.2–2 GHz. Each CF shifts by half of this change, with opposite signs for the two branches. The geometric and photoelastic contributions from Eqs. (S46) and (S50) partly cancel, giving a much smaller common midpoint shift of approximately −1 MHz for 1 nm squeezing and −30 MHz for 10 nm squeezing. These estimates therefore indicate that squeezing by 1–10 nm can induce CF shifts of approximately 0.1–1 GHz, comparable to the observed discrepancy. The numerical values are approximate and depend on the simplified local contact model.